\documentclass[aps,9pt,notitlepage]{revtex4-1}
\usepackage[utf8]{inputenc}

\usepackage{subcaption}
\usepackage{graphicx}

\usepackage{verbatim}

\usepackage{physics}
\usepackage{amssymb}
\usepackage{bm}
\usepackage{braket}
\usepackage{stackengine}
\usepackage{siunitx}
\let\u=\SI%

\newcommand{\Chii}{\raisebox{2pt}{$\chi^{(2)}$}}

\newcommand{\wi}{\omega_1}
\newcommand{\ws}{\omega_2}
\newcommand{\wpump}{\omega_p}
\newcommand{\wcp}{\omega_{cp}}
\newcommand{\qi}{\mathbf{q}_1}
\newcommand{\qone}{\mathbf{q}_2}
\newcommand{\q}{\mathbf{q}}
\newcommand{\qs}{\mathbf{q}_2}
\newcommand{\qtwo}{\mathbf{q}_2}
\newcommand{\qp}{\mathbf{q}_p}

\newcommand{\ki}{\mathbf{k}_1}
\newcommand{\ks}{\mathbf{k}_2}
\newcommand{\G}{G^{(1)}}
\newcommand{\GG}{G^{(2)}}
\newcommand{\ro}{\boldsymbol{\rho}}
\newcommand{\rhoi}{\boldsymbol{\rho}_1}
\newcommand{\rhos}{\boldsymbol{\rho}_2}

\newcommand\Tstrut{\rule{0pt}{2.6ex}}         

\DeclareMathOperator{\sinc}{sinc}

\begin{document}
\title{Characterization of space-momentum entangled photon with a time resolving CMOS SPAD array}

\author{Bruno Eckmann$^1$, Bänz Bessire$^1$, Manuel Unternährer$^1$, Leonardo Gasparini$^2$, Matteo Perenzoni$^2$, André Stefanov$^1$}

\affiliation{$^1$ Institute of Applied Physics, University of Bern, 3012 Bern, Switzerland}
\affiliation{$^2$ Fondazione Bruno Kessler FBK, 38122 Trento, Italy}

\vspace{10pt}

\begin{abstract}
Single photon avalanche diode arrays can provide both the spatial and temporal information of each detected photon. We present here the characterization of entangled light with a sensor specifically designed for quantum imaging applications. The sensors is time-tagging each detection events at the pixel level with sub-nanosecond accuracy, within frames of 50 ns. The spatial correlations between any number of detections in a defined temporal window can thus be directly extracted from the data. We show the ability of the sensor to characterize space-momentum entangled photon pairs emitted by spontaneous parametric downconversion. Their entanglement is demonstrated by violating an EPR-type inequality. 
\end{abstract}

\maketitle

\section{Introduction}

Quantum states of light are a fundamental tool to implement quantum information processing protocols and quantum metrology methods in the optical domain, as quantum imaging \cite{Genovese2016}. They can be described either by discrete variables, the photons, or by continuous fields. Correspondingly measurements of light can be mainly split into two classes: measurements of fields, by homodyne measurements for example, and intensity measurements, that correspond fundamentally to photon counting. These are the tool of choice to characterize discrete quantum optical states; for example to measure correlations in quantum states with few photons. An ideal detector should be able to detect the position and time of arrival of all impinging photons. Such universal detector do not yet exist, but there are various type of single photon sensitive detectors, that are optimized with respect to specific characteristics. Most of the experiments based on correlation between photons are realized with single photon detectors with high temporal resolution but no spatial resolution \cite{Eisaman2011a}, including photomultipliers, single photon avalanche diodes (SPAD) or superconductor detectors. On the other side, imaging experiments are usually implemented, even at the single photon level, with low noise cameras as EMCCD or scientific CMOS cameras \cite{Seitz2011,Moreau2019}, that are able to localize single photons among many pixels, but have low temporal resolution. Spatial entanglement in photon pairs have been demonstrated with EMCCD \cite{Edgar2012}\\
A way to combine high temporal and spatial resolution is fast gating of imaging sensors. Detection of spatial entangled photon pairs have been realized with ICCD cameras as in \cite{Fickler2013} and commercial gated single photon avalanche array sensors \cite{Ndagano2020}. However, while gating gives access to short times, the full spatio-temporal correlations of light across a wide range of detection times cannot be measured. In addition, gating is well suited for pulsed operations, but a continuous source of correlated photons would ideally be measured with a non-synchronous detector. This is easily implemented with single pixel detectors, that are always ready to trigger, up to their deadtime. However when operating many detectors in parallel, as in detector arrays, the readout mechanism usually requires a frame-based operation. In that case one of the criterion to be optimized is the sensor duty cycle, given by the frame rate of the detector times the duration of each frame. For gated detectors, the requirement to have short gates to improve the temporal discrimination of the correlated photons contradicts the need of long frames to improve the photons collection efficiency. The solution is to be able time-stamping with high temporal resolution each detection event within a long frame. Combining single photon detectors and external time-stamping electronics can be realized only for a limited number of detectors \cite{Tenne2019}. This cannot however be extended to imaging with thousands of pixels without integrating light sensitive devices with digital electronics. CMOS offers the possibility of full integration for silicon sensors. Here we demonstrate the capability of a recently developed SPAD array sensor \cite{Gasparini2018,Unternahrer2017a} to characterize spatial entanglement from spontaneous parametric downconversion (SPDC) by measuring both near- and far-field correlations \cite{Howell2004,Edgar2012}. We demonstrate entanglement of the state by testing an Einstein Podolsky Rosen (EPR)-type inequality. We estimate the violation of the inequality both  directly from the measured join probabilities, without a priori assumption on the quantum state, and from fitted data assuming a gaussian model for the SPDC emission. \\
In section \ref{sec:Theory} we recall the quantum state of the photon pairs emitted by SPDC and of the expected join probability distribution that have to be measured in order to verify their spatial entanglement, by violating an EPR-type inequality. Section \ref{sec:Experiment} describes the experimental setup and the processing of the data acquired with the SPAD array with corrections of the accidental events and cross-talk. Finally, the results are presented in section \ref{sec:Results}, where the different estimations of the violation of the inequality are presented.

\section{Theory of spatial entanglement in SPDC} \label{sec:Theory}
Spontaneous Parametric Down-Conversion (SPDC) is nowadays a common source of entangled photon pair, where a pump photon of frequency $\wpump$ is annihilated inside a non-linear crystal (NLC) with a non-vanishing second order susceptibility $\chi^{(2)}$ and two photons with frequencies $\wi$ and $\ws$ and wavevectors $\ki$ and $\ks$ (historically called \emph{signal} and \emph{idler}) are simultaneously created \cite{Couteau2018}. Here we consider the case of type-0 quasi-phase-matching in a periodically poled crystal of length $L$ with poling period $G$ and refractive index $n(\omega)$, where all fields have the same polarization.

Under the approximation of a monochromatic pump  with a frequency $\omega_{cp}$, the envelop of the pump field is given by 
\begin{equation}
\mathcal{E}_p^+(\qp,\wpump) = 2\pi \mathcal{E}_p^+(\qp) \delta(\wpump-\wcp).
\end{equation}
The quantum state of the generated signal and idler photons is then given by 

\begin{equation}
\ket{\Psi} = \ket{0} + \int d^2 q_i d^2 q_s d\ws \ \Lambda(\qi, \qs, \omega_s) \ket{\qi, \wcp-\ws}\ket{\qs,\ws},
\label{eq:spdcstate_mono}
\end{equation}
with the one photon Fock states $\ket{\mathbf{q}_j,\omega_j} =\hat{a}^\dagger(\mathbf{q}_{j},\omega_{j}) \ket{0}_j$ and where $\q_{i}=(k_{x_i}, k_{y_i})$ is the transverse component of the wave vector $\mathbf{k}_i$.  The two-photon wavefunction, or joint spectral amplitude (JSA) is
\begin{equation}
\begin{split}
\Lambda(\qi, \qs, \omega_s) = &-\frac{4i\epsilon_0\Chii L e(\wcp-\ws)e(\ws)}{3\hbar\pi(2\pi)^5\ n(\wcp-\ws)n(\ws)}\mathcal{E}_p^+(\qi+\qs)\\
&\times \sinc\left(\frac{\Delta k_z L}{2}\right)\   \exp\left(-i\frac{\Delta k_z L}{2}\right ),
\end{split}
\label{eq:jsamono}
\end{equation}
with the phase matching relation is
\begin{equation}
\label{eq:deltakzmono}
\begin{split}
\Delta k_{z} = &\sqrt{\left (\frac{\wcp-\ws}{c}n(\wcp-\ws)\right )^2-\qi^2} + \sqrt{\left (\frac{\ws}{c}n(\ws)\right )^2-\qs^2}\\
&- \sqrt{\left (\frac{\wcp}{c}n(\wcp)\right )^2-(\qi+\qs)^2} + \frac{2\pi}{G},
\end{split}
\end{equation}
and the normalization function
\begin{equation}
e(\omega) = i\sqrt{\frac{\hbar\omega}{2(2\pi)^3\epsilon_0 c}}.
\end{equation}

We note that the JSA $\Lambda(\qi,\qs, \ws)$ given by \eqref{eq:jsamono} cannot in general be factorized  into functions $\Lambda_i(\qi, \ws)$ and $\Lambda_s(\qs, \ws)$ and thereby the state is entangled. Assuming narrow-band frequency filtering of the down-converted photons and thus only considering the spatial dependency, the JSA factorizes into momentum and frequency parts \cite{Unternahrer2018}
\begin{equation}
\Lambda(\qi,\qs, \omega_s)=\Lambda(\qi,\qs)S(\ws)
\end{equation}
Equivalently the JSA can be expressed in term of transverse positions $(\rhoi,\rhos)$ and times $(t_1,t_2)$ through Fourier transform $\mathcal{F}$
\begin{equation}
\tilde{\Lambda}(\rhoi,\rhos)\tilde{S}(\tau)=\mathcal{F}\{\Lambda(\qi,\qs)\}\mathcal{F}\{S(\ws)\}
\end{equation}

In the monochromatic approximation, the temporal part only depends on the time difference $\tau=t_1-t_2$. Coincidence detections measure events around $\tau=0$. Therefore the temporal part of the JSA only account for a proportionality factor and thus will be further omitted.

The position $\boldsymbol{\rho}_i=(x_i,y_i)$ and momentum $\hbar\boldsymbol{q}_i=(\hbar k_{x_i}, \hbar k_{y_i})$ operators do not commute and therefore a criterion for EPR-type correlations between two systems $1$ and $2$ (that correspond in our case to the signal and idler photons) can be established \cite{Reid2009}. In the SPDC emission, the $x$ and $y$ components are decoupled. We thus consider the one dimensional case with $x_i$ and $p_i=\hbar k_{x_i}$. The results of correlation measurements are therefore described by the join probabilities $\mathcal{P}(\boldsymbol{\rho}_1,\boldsymbol{\rho}_2)$ and $\mathcal{P}(\boldsymbol{q}_1,\boldsymbol{q}_2)$, or $\mathcal{P}(x_1,x_2)$ and $\mathcal{P}(p_1,p_2)$ in the 1D case. The minimal inferred variance of $(x_1,x_2)$ is defined by
\begin{equation}
\Delta_{min}^2(x_1|x_2) = \int dx_s\ \mathcal{P}(x_2) \Delta^2(x_1|x_2),
\label{eq:epr_minimuminferredvariance_x}
\end{equation}
where $\Delta^2(x_1|x_2)$ is the variance of the conditional probability $\mathcal{P}(x_1|x_2)$ and $\mathcal{P}(x_2)$ is the marginal probability of system $2$. The same definitions apply for the variables $(p_1,p_2)$. 
According to \cite{Reid2009}, the fulfillment of the following inequality on the product of minimum inferred variances indicates the EPR-type correlations 
\begin{equation}
\Delta_{min}^2(x_1|x_2)\Delta_{min}^2(p_1|p_2)< \frac{\hbar^2}{4},
\end{equation}
or expressed as a dimensionless quantity
\begin{equation}\label{eq:EPRInequality}
\Delta_{min}^2(x_1|x_2)\Delta_{min}^2(k_{x_1}|k_{x_2})< \frac{1}{4}.
\end{equation}
Experimentally the join probabilities are derived from correlation measurements, that correspond to second order coherence functions of the electric field. They are given at transverse positions $\ro_1$ and $\ro_2$ and times $t_1$ and $t_2$ by
\begin{equation}\label{eq:gtwo}
\GG(\ro_1,t_1;\ro_2,t_2) = \ev{\hat{E}^-(\ro_1,t_1)\hat{E}^-(\ro_2,t_2)\hat{E}^+(\ro_2,t_2)\hat{E}^+(\ro_1,t_1)}{\Psi}.
\end{equation}
Correspondingly, in the monochromatic approximation the join probability is directly proportional to the second order coherence function 
\begin{equation}
\mathcal{P}(\boldsymbol{\rho}_1,\boldsymbol{\rho}_2) \propto \GG(\boldsymbol{\rho}_1,\boldsymbol{\rho}_2)
\end{equation}
and is explicitly related at the crystal position to the spatial JSA of the state \eqref{eq:spdcstate_mono} by
\begin{equation}\label{eq:gtwo_spdc}
\GG(\boldsymbol{\rho}_1,\boldsymbol{\rho}_2) \propto \abs{\tilde{\Lambda}(\boldsymbol{\rho}_1,\boldsymbol{\rho}_2) }^2.
\end{equation}
In general, the imaging setup from the crystal to the sensor plan determines the field operators that have to be introduced in \eqref{eq:gtwo}. In the following we performed the two types of measurements relevant for inequality (\ref{eq:EPRInequality}): near- and far-field imaging. 

\subsection{Second Order Near-Field Correlations}
When imaging the crystal with a magnification $M$, the correlation function at the image plan using Fourier optics can be expressed as

\begin{equation}\label{eq:nearfieldcorrelations}
\GG(\ro_1,\ro_2) \propto \abs{\int d^2 q_i \ \int d^2 q_s  \Lambda(\qone, \qtwo) e^{i(\qone\ro_1 + \qtwo\ro_2)/M} }^2. 
\end{equation}
This is essentially the Fourier transform of the propagated two-photon wave function.

\subsection{Second Order Far-Field Correlations}
In a far-field imaging setup, using a lens of focal length $f$, the relation between the transverse position at the imaging plan $\ro$ and the transverse momentum $\q$ at the object is 
\begin{equation}
\q = \frac{k}{f}\ro,
\end{equation}
where $k$ is the light wave-vector.
Hence, the second-order correlation function in the far-field reads 
\begin{equation}
\GG(\ro_1,\ro_2) \propto \abs{\Lambda(\frac{k}{f}\ro_1,\frac{k}{f}\ro_2)}^2.
\end{equation}
The spatial dependency of the correlation function thus directly reflects the JSA expressed in the momentum space.

\section{Experimental evaluation of correlations} \label{sec:Experiment} 
The join probability distribution that leads to the evaluation of Eq. \eqref{eq:EPRInequality} has to be estimated from the raw correlations acquired with the sensor. Ideally it should be estimated directly from the non-processed data, in order to achieve a non-conditional violation of the inequality. However due to the sensor imperfections: reduced efficiency, dark counts and cross-talk between pixels, some assumptions have to be introduced in order to process and correct the raw data. In a second step, the corrected data can be either directly used to numerically evaluate Eq.\eqref{eq:EPRInequality} or, by introducing additional assumptions, fitted with a model of the expected correlations as for instance in \cite{Edgar2012}.

\subsection{Experimental setup}
\begin{figure}[h]
	\centering
	\includegraphics[width=1\columnwidth]{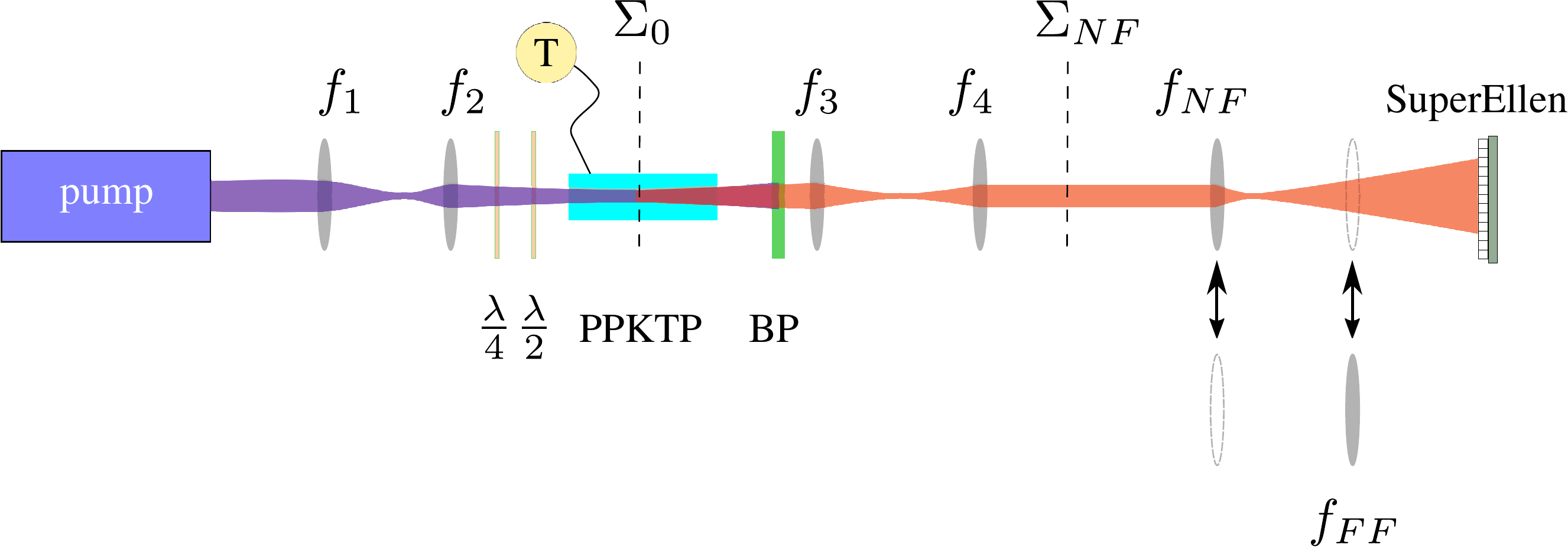}
	\caption{Experimental setup for spatial correlation measurements. A CW laser at \u{405}{\nano\meter} is slightly focused by a telescope ($f_1$ and $f_2$) and pumps the crystal. The bandpass filter (BP) only transmits the down-converted photons at a central frequency of \u{810}{\nano\meter}. The 4-f lens configuration ($f_2$ and $f_3$) one to one images the crystal center $\Sigma_0$ onto the near field plane $\Sigma_{NF}$. Further imaging onto the detector array occurs by either a 1 lens configuration $f_{NF}$ for the near field or a fourier lens $f_{FF}$ for the far field.}
	\label{fig:experiment_setup}
\end{figure}
The experimental setup is shown in Fig.\ref{fig:experiment_setup}. A continuous wave (CW) laser (Toptica DL PRO HP 405) with a maximal power output of \u{30}{\milli\watt} at a wavelength of \u{405}{\nano\meter} and a spectral bandwidth of \u{80}{\mega\hertz} serves as the pump.  It is slightly focuses onto the crystal's center plane $\Sigma_0$ by a two lens system ($f_1=f_2=\u{200}{\milli\meter}$). The beam waist of the pump at this plane is $w_{0x}=\u{250}{\micro\meter}$ in the $x$ direction and and $w_{0y}=\u{300}{\micro\meter}$ in the $y$ direction. $\lambda/4$- and a $\lambda/2$-plates are used to achieve the desired horizontal polarization in the crystal. A $1\times2\times\u{12}{\milli\meter\cubed}$ periodically poled KTiOPO$_4$ (PPKTP) non-linear crystal with poling period $G_0=\u{3.51043}{\micro\meter}$ is embedded into a temperature controlled oven and provides the source for the down-converted photons. The oven is maintained at a temperature of \u{26.0}{\degreeCelsius} for an almost collinear phase-matching, that maximize the photon flux onto the sensor in the far field. The down-converted photons are separated from the pump by a bandpass filter (BP) centered at \u{810}{\nano\meter} and with FWHM \u{10}{\nano\meter}. Given the source parameters, waists of the pump beam and the length of the SPDC crystal, we can estimate the minimal inferred standard deviations to be

\begin{align}
	\Delta_{min}(x_1|x_2) \approx \u{37.3}{\micro\meter} \qquad \Delta_{min}(k_{x_1}|k_{x_2}) \approx  \u{4.0}{\per\milli\meter},\\	\Delta_{min}(y_1|y_2) \approx \u{37.3}{\micro\meter} \qquad \Delta_{min}(k_{y_1}|k_{y_2}) \approx  \u{3.4}{\per\milli\meter}.
\end{align}
and the violation of the Heisenberg-inferred inequality to be
\begin{align}
	V_{min}^{(x)} &\equiv \Delta^2_{min}(x_1|x_2)\cdot \Delta^2_{min}(k_{x_1}|k_{x_2}) \approx \u{2.2e-2}{}\\
	V_{min}^{(y)} &\equiv \Delta^2_{min}(y_1|y_2)\cdot \Delta^2_{min}(k_{y_1}|k_{y_2}) \approx \u{1.6e-2}{}.
\end{align}

Two lenses ($f_3=f_4=\u{50}{\milli\meter}$) form a $4f$ imaging system such that the electric field at the plane $\Sigma_{NF}$ is an exact replica of the field in the crystal at $\Sigma_{0}$. Either the near- or far-field of this plan is then imaged onto the sensor, depending on the selected lens. For near-field, a lens $f_{NF}=\u{25.4}{\milli\meter}$ images the plane $\Sigma_{NF}$ onto the sensor with a magnification factor $M=9$. For far-field, a lense $f_{FF}=\u{150}{\milli\meter}$ images the far-field such that the area covered by the sensor corresponds in $k$-space to $\pm\u{36.2}{\per\milli\meter}$, as $q = x \frac{k}{f_{FF}}$.

The coincidence detection is performed with a single photon avalanche detector (SPAD) array recently developed \cite{Gasparini2018}. It is a fully digital $32\times32$ pixels sensor array based on CMOS technology. The total sensitive area is $1.4\times1.4 \text{ mm}^2$ and the pixel pitch $\Delta L$ is $\u{44.67}{\micro\meter}$ with an overall fill-factor of $\u{19.48}{\percent}$. A Time-to-digital converter (TDC) integrated in each pixel allows for pixel-wise time stamping the first detection event within a frame. The time resolution (time-bin length) of the TDC is $\u{205}{\pico\second}$ and the frame length is 255 time-bins that corresponds to about $\u{50}{\nano\second}$. A maximal frame observation rate of $\sim\u{1}{\mega\hertz}$ can be reached, which corresponds to a duty cycle of $\u{4.5}{\percent}$ . The total photon detection efficiency is $\u{5}{\percent}$ at $\u{400}{\nano\meter}$ and $\u{0.8}{\percent}$ at $\u{810}{\nano\meter}$. The dark count rate is below \u{1}{\kilo\hertz} at room temperature on average. The data from the sensor's FPGA are transfered to a computer through a USB 3.0 interface. A Labview interface controls the acquisition, allowing for either on-line simple processing of the data, or acquisition of all raw sensor data. For the following measurements, we acquired the raw data and extracted the correlations off-line, with Matlab and C programs. 

\subsection{Data processing}

\begin{figure}[h]
	\centering
	\includegraphics[width=0.5\columnwidth]{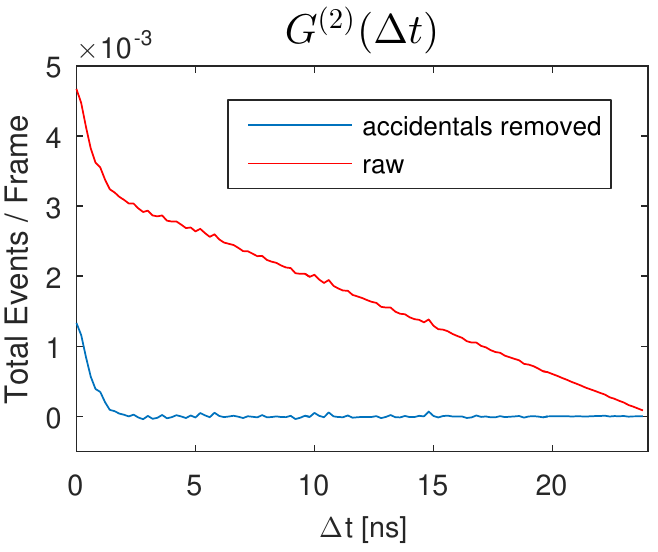}
	\caption{The temporal second-order correlation $G^{(2)}(\Delta t)$ accumulated over all pixels. The detection event counts are shown in raw form and after removal of uncorrelated accidentals.}
	\label{fig:G2_dt}
\end{figure}

The coordinate and time of all detection events within one frame, together with the frame identification number, constitute the acquired raw data. In the following we associate the spatial coordinate $\boldsymbol{\rho}_i=(x_i,y_i)$ with the pixel coordinate $\mathbf{p}_{i}=(\mathrm{p}_{x_{i}},\mathrm{p}_{y_{i}})$ using $\ro_{i}=\Delta L\mathbf{p}_{i}$ . An estimation of the correlation $\GG(\mathbf{p}_1,\mathbf{p}_2)$ between a pair of pixels is obtained by counting all coincidence events between those pixels that occur within a defined temporal window. The histogram of all time differences between pairs of detection events across the sensor is shown on Fig. \ref{fig:G2_dt}. While the temporal jitter of each pixel is of the order of 200 ps, in practice temporal shifts across the sensor widen the coincidence peak. In principle, a pixel-wise temporal calibration would allow to narrow the window. Here we define the coincidence window to be $10$ TDC steps ($\sim \u{2}{\nano\second}$) in order to catch all coincidence events.  The obtained histogram is further normalized to a number of coincidence per frames or million of frames (MFrames) \cite{Unternahrer2018}. The second order correlation in a transverse plane is a function of four variables $\GG(\mathbf{p}_1,\mathbf{p}_2)=\GG(\mathrm{p}_{x_1},\mathrm{p}_{y_1}, \mathrm{p}_{x_2}, \mathrm{p}_{y_2})$.

\begin{figure}[htb]
	\topinset{\includegraphics[height=4cm]{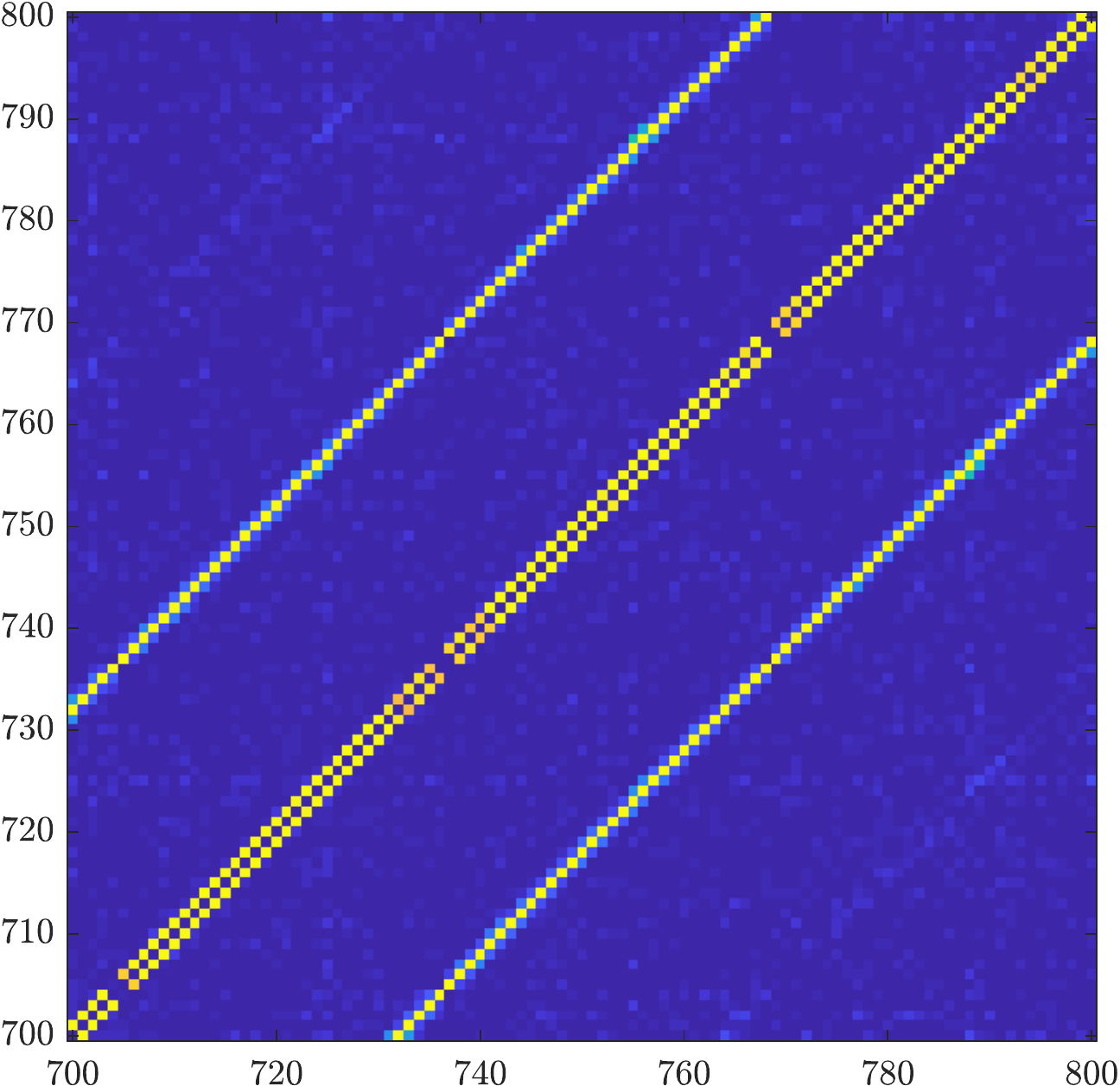}}
	{\includegraphics[width=1\linewidth]{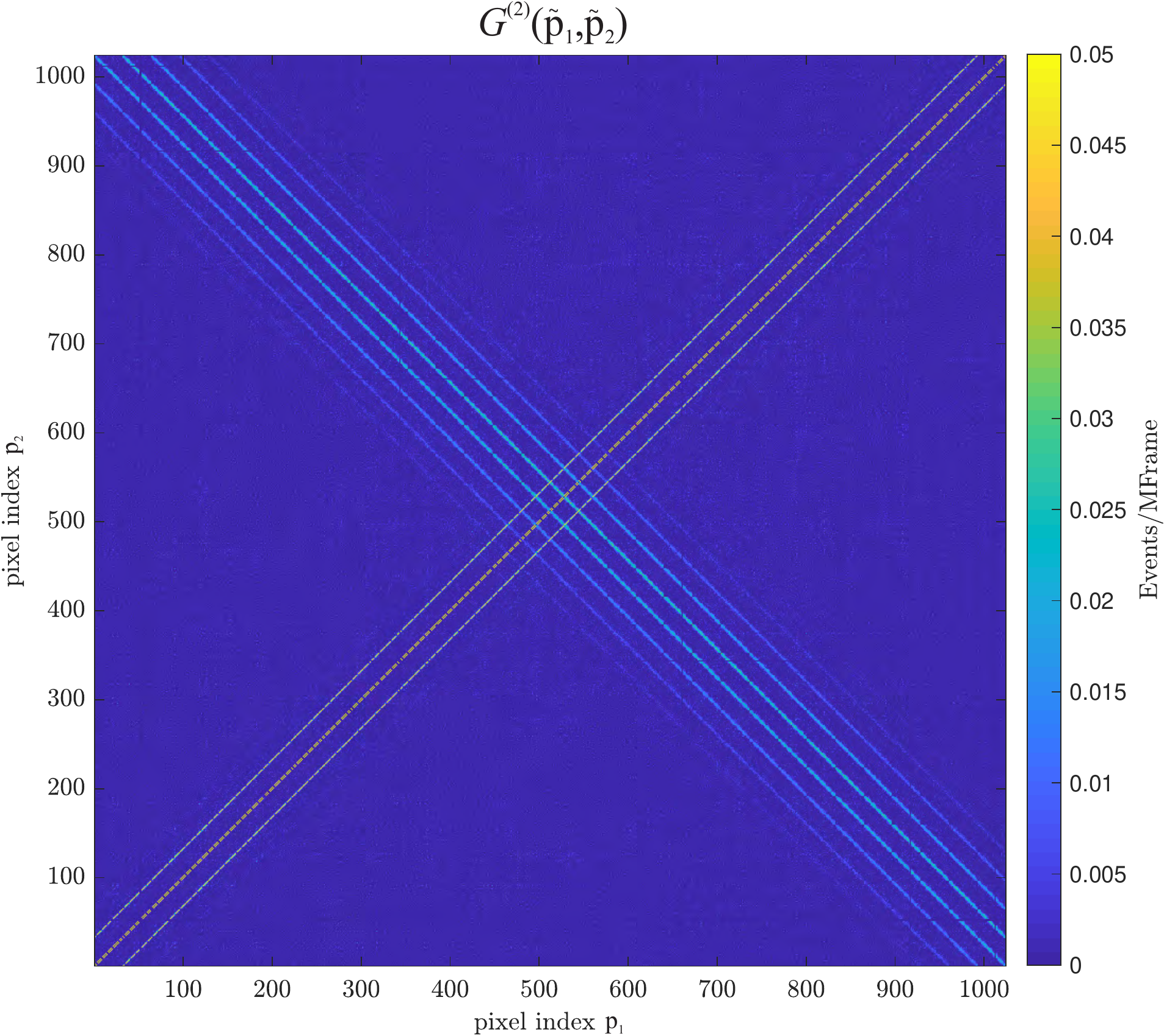}}{20pt}{5pt}
	\caption{Full second order correlations between every pixel pairs in the far-field, with corrected accidentals. Each pixel is addressed with a linear pixel index $\tilde{\mathrm{p}}_{i}=1,\dots,1024$. The inset shows a close up of the cross-talk correlations appearing along the diagonals. }
	\label{fig:g2_full}
\end{figure}

In order to be able to plot a 2D representation of the full correlations, we introduce the pixel indexes numerating each pixel from $1\dots1024$
\begin{equation}
\tilde{\mathrm{p}}_{i} = \mathrm{p}_{x_{i}} + 32\cdot(\mathrm{p}_{y_{i}}-1).
\end{equation}
Figure \ref{fig:g2_full} shows the full correlation matrix between every pair of pixels after accidentals subtraction (see bellow) in the case of far-field measurement. The photon pairs exhibit anti-correlation in their detection position, that reflects into the $\approx 8$ anti-diagonal lines. The exact diagonal cancels, as the sensor cannot measure self-correlations on one pixel. The nearest neighbors correlations (separated by pixel indexes $\pm1$ and $\pm 32$) are affected from cross-talk, forming the four diagonal lines.

In order to apply the inequality \eqref{eq:EPRInequality} for only one dimension, the correlation functions are projected onto either $x$ or $y$ dimnesions by
\begin{align}
	\GG(x_1,x_2) &= \sum_{y_1} \sum_{y_2} \GG(x_1,y_1,x_2,y_2),\\
	\GG(y_1,y_2) &= \sum_{x_1} \sum_{x_2} \GG(x_1,y_1,x_2,y_2).
\end{align}
Alternatively, one can use the centroid and difference coordinates to project the 4D correlation function on 2D plans defined by fixed value of $\ro_1+\ro_2$ or $\ro_1-\ro_2$, respectively. The projected correlation functions read
\begin{align}
	\GG(\ro_+) &= \sum_{\substack{\ro_1,\ro_2\\ \ro_1+\ro_2 = \sqrt{2}\ro_+}}  \GG(\ro_1,\ro_2),\\
	\GG(\ro_-) &=  \sum_{\substack{\ro_1,\ro_2\\ \ro_1-\ro_2 = \sqrt{2}\ro_-}}\GG(\ro_1,\ro_2).
\end{align}.

\subsubsection{Removing Accidentals}
Accidental coincidences are coincidences events that occur in the defined coincidence window but are neither temporally nor spatially correlated. They steam from detections triggered by background light, dark counts, and mostly from SPDC photons that do not belong to the same photon pair. The total measured raw correlation signal is thus given by
\begin{equation}
\GG_{raw}(\ro_1,\ro_2) = \GG_{corr}(\ro_1,\ro_2) + \GG_{acc}(\ro_1,\ro_2).
\end{equation}

\begin{figure}[htb]
	\begin{subfigure}[b]{0.3\linewidth}
		\includegraphics[width=1\linewidth]{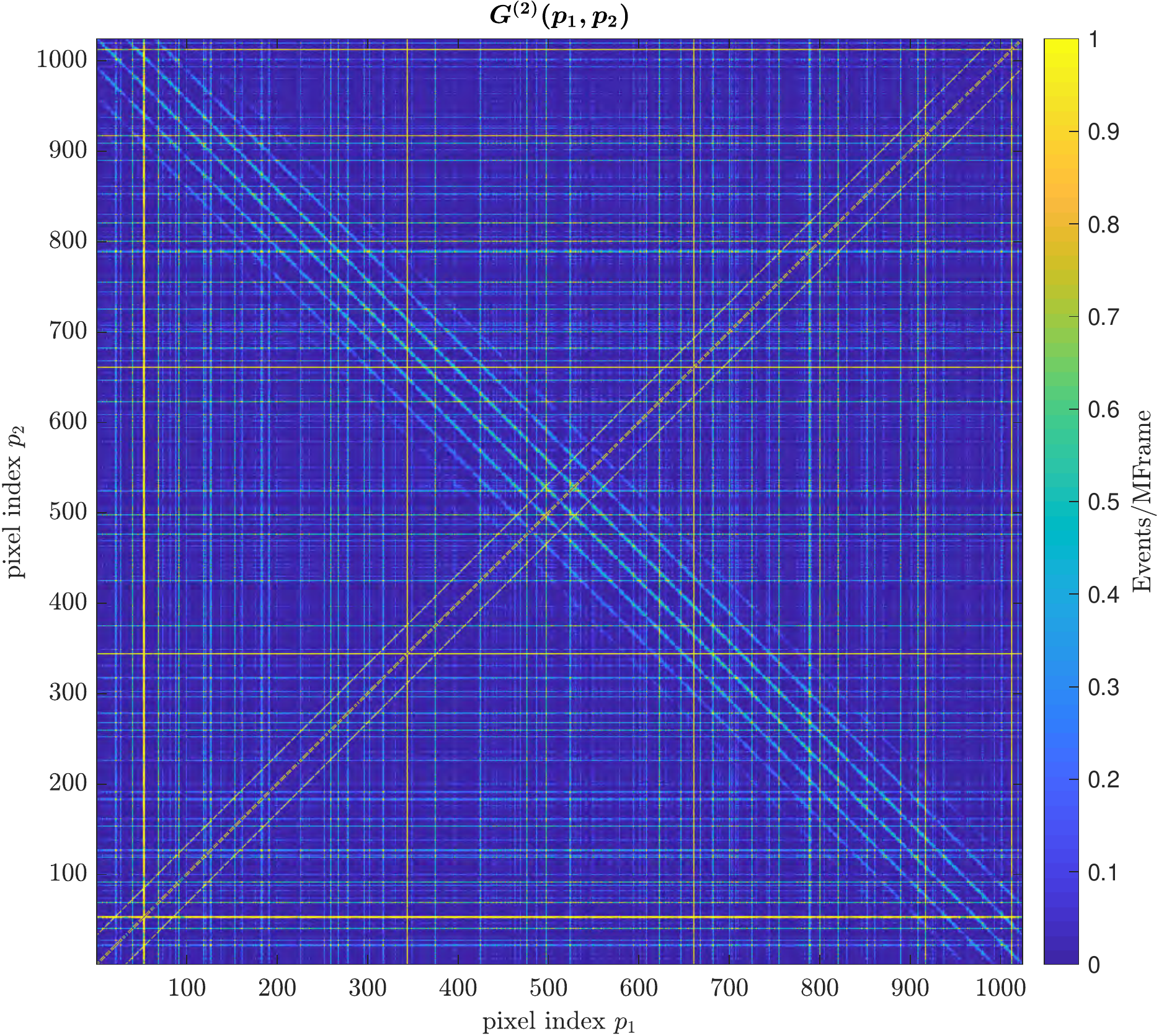}
		\subcaption{}
	\end{subfigure}
	\begin{subfigure}[b]{0.3\linewidth}
		\includegraphics[width=1\linewidth]{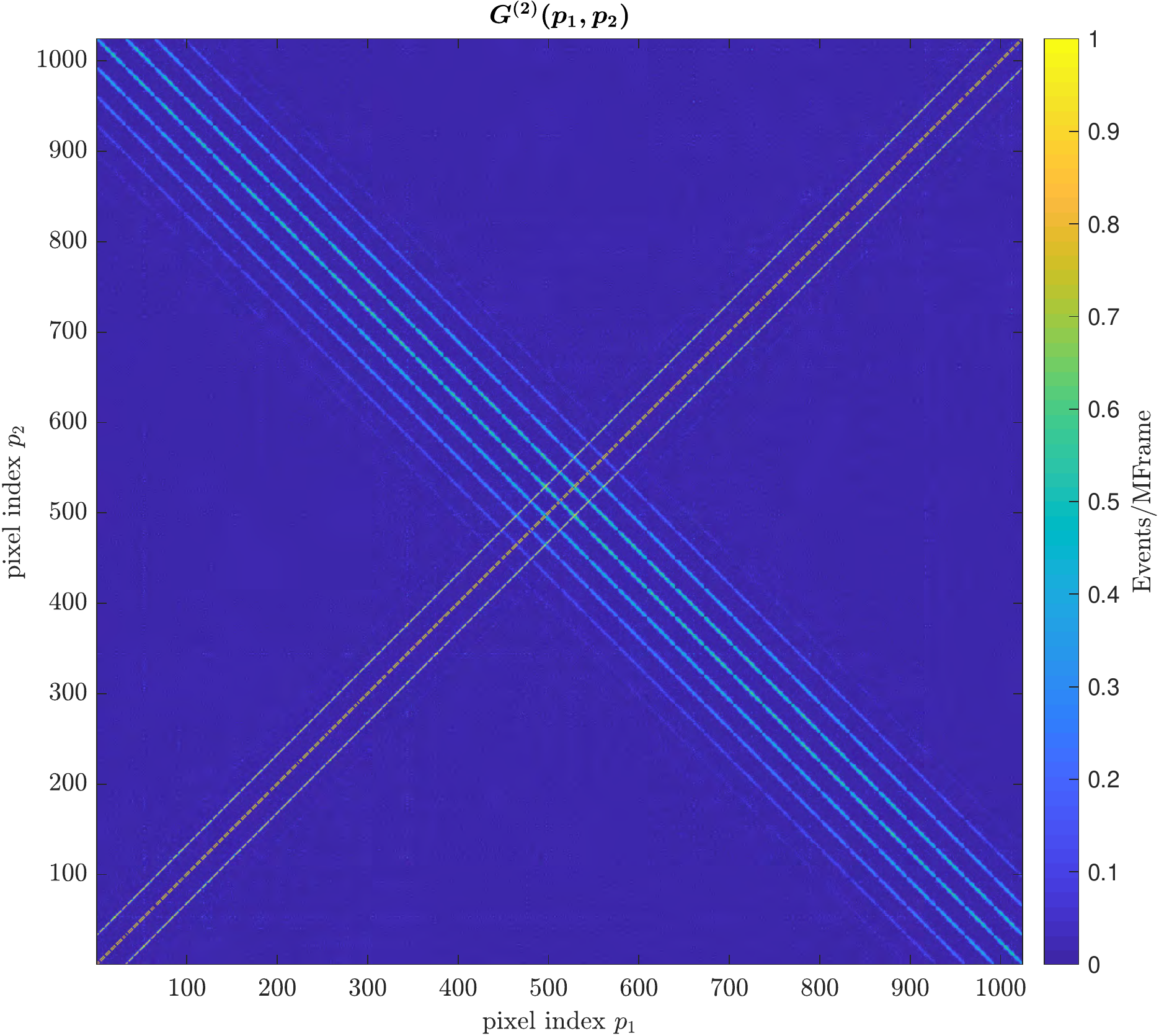}
		\subcaption{}
	\end{subfigure}
	\begin{subfigure}[b]{0.3\linewidth}
		\includegraphics[width=1.05\linewidth]{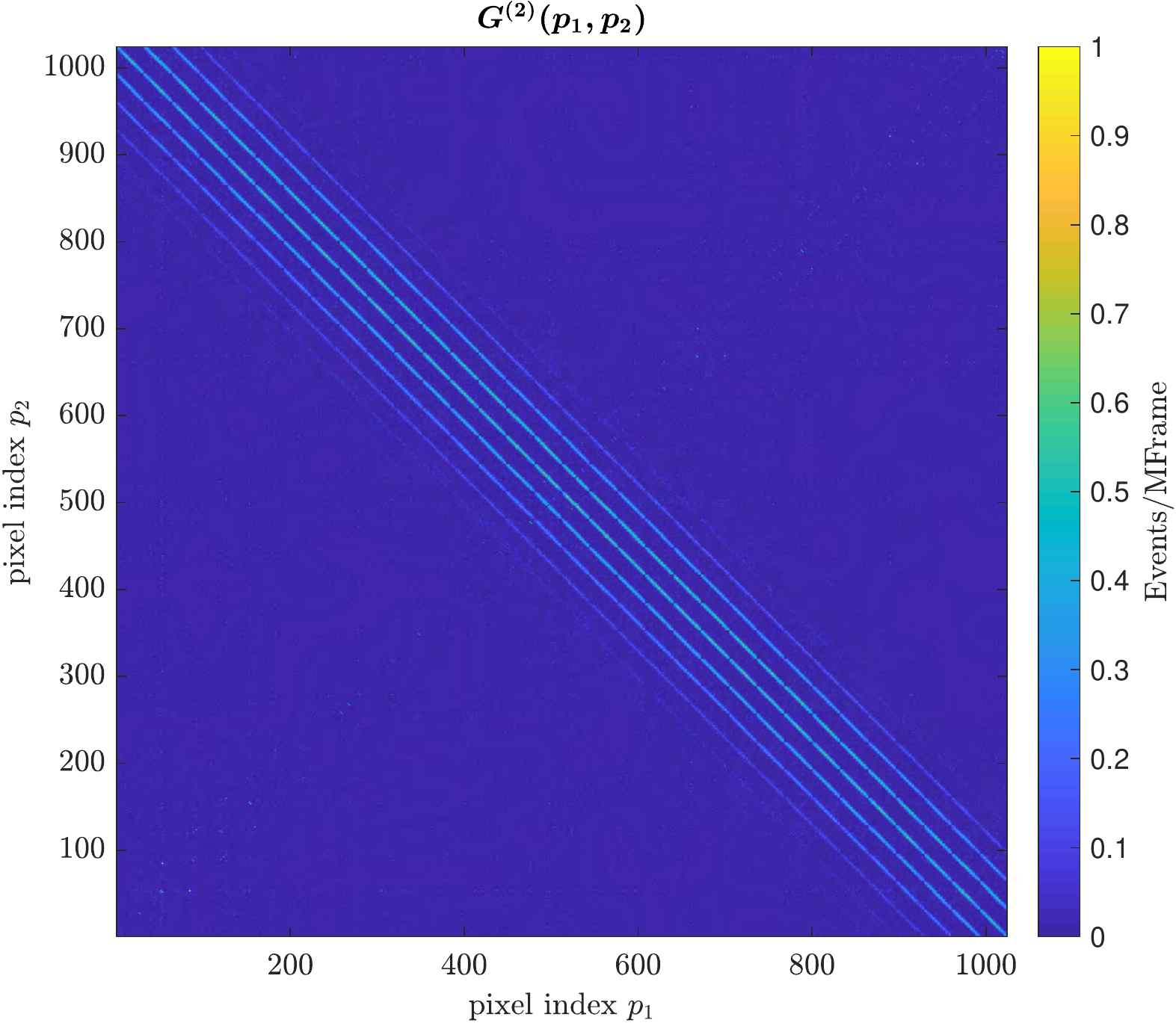}
		\subcaption{}
	\end{subfigure}
	\caption{Second order far-field correlations. Measured raw data (a), after removing accidentals (b) and after cross-talk correction (c)}
	\label{fig:g2full_accidentals}
\end{figure}

These accidentals events contributes significantly to the raw signal as we can observe on Fig. \ref{fig:g2full_accidentals} (a). They can be however estimated and corrected for, either by measuring the coincidence signal in a shifted time window or by using the fact that uncorrelated light obeys
\begin{equation}
\GG_{acc}(\ro_1,\ro_2) \propto \G(\ro_1) \otimes \G(\ro_2).
\end{equation}
The proportionally factor depends on the coincidence window length, and is estimated by comparing event rates in $\GG(\ro_1,\ro_2)$ and  $\G(\ro_1) \otimes \G(\ro_2)$ for pixels for which no correlations are expected. $\G(\ro)$ is the first order correlation function, or corresponding to the intensity. Figure \ref{fig:g2full_accidentals} shows the raw (a) and corrected data after removing accidental coincidences (b). The anti-correlations from the photon pairs is clearly visible, but also correlations due to cross talk, that are further removed in Figure \ref{fig:g2full_accidentals} (c) (see bellow).

\subsubsection{Cross-talk}
\begin{figure}[htb]
	\begin{subfigure}[b]{0.45\linewidth}
		\includegraphics[width=0.8\linewidth]{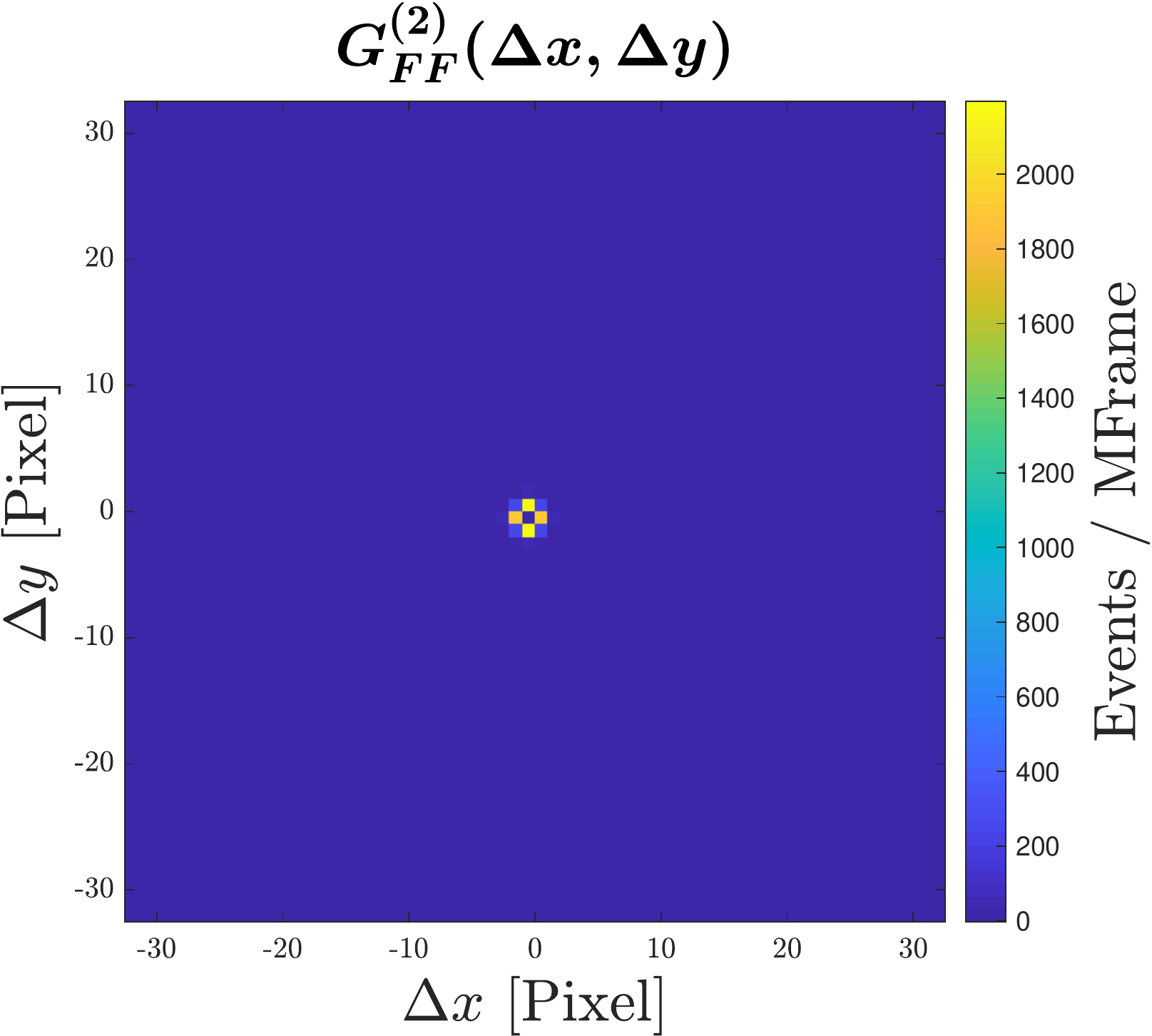}
		\subcaption{}
	\end{subfigure}
	\begin{subfigure}[b]{0.45\linewidth}
		\includegraphics[width=0.8\linewidth]{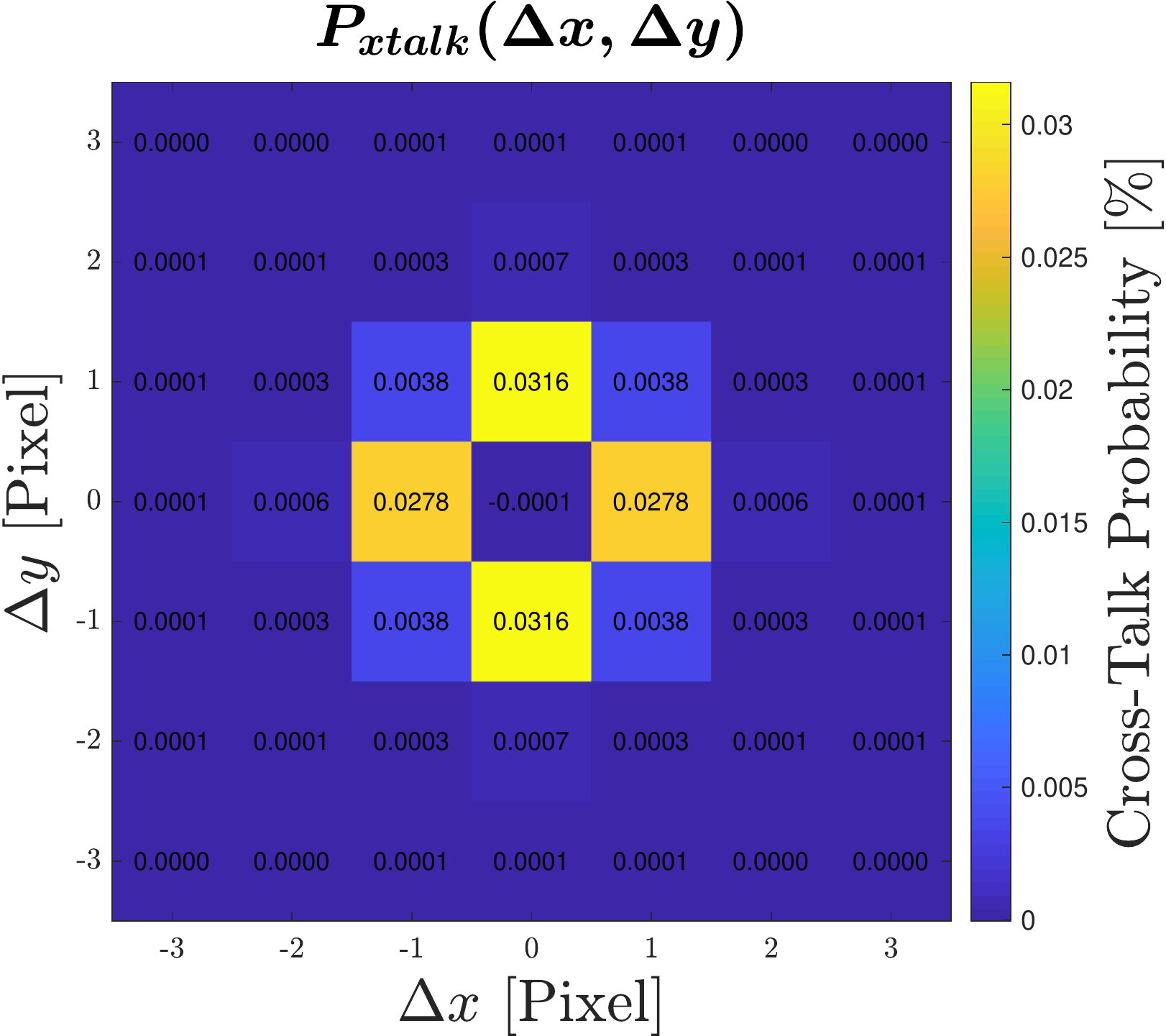}
		\subcaption{}
	\end{subfigure}
	\caption{(a) Correlation-peak due to cross-talk of the second order correlation measurement in the far-field. (b) Cross-talk probability extracted from the correlation peak and normalized with the total number of counts. The pixel at $(\Delta x, \Delta y)=0$ is the emitter of the cross-talk.}
	\label{fig:crosstalk}
\end{figure}

A single detection event, triggered by a dark count or a photon can trigger nearby pixels. This leads to undesirable detection events called \text{cross-talk} and are an artifact intrinsic to the detector. The physical process behind \textit{optical crosstalk} are photons, created by the charge avalanche of the first triggering SPAD. These photons may reach and trigger neighboring  pixels, in a very short time scale.
Cross-talk is especially undesired in the near-field correlations, as it overlays the real signal. The cross-talk probability, averaged over all pixels, can be extracted from the correlation-peak in the far-field, as in that case only anti-correlations are expected. Hence, the mean probability that a pixel at distance $(\Delta{x},\Delta{y})$ is triggered from cross-talk is given by
\begin{equation}
\label{eq:crosstalkprob}
P_{xtalk}(\Delta x,\Delta y) = \frac{1}{2} \frac{\GG_{FF}(\Delta {x},\Delta {y})}{\sum_x\sum_y \G_{FF}(x,y)},
\end{equation}
where the sum over $\G_{FF}$ normalizes with the total number of counts. To take into account that we cannot measure all cross-talk events of the pixels close to the sensor's edges, only an inner window of the total sensor area is used to estimate the cross-talk behavior. $\GG_{FF}$ in \eqref{eq:crosstalkprob} is reduced to cover only an inner window of $29\times29$ pixels. Consequently, the normalization is also adjusted to the total counts within this window. The second order correlation measurements can then be corrected \cite{Lubin2019} according to
\begin{equation}
\begin{split}
\GG(x_1,y_1,x_2,y_2) = &\GG_{m}(x_1,y_1,x_2,y_2)\\
&- P_{xtalk}(x_2-x_1,y_2-y_1) \G(x_1,y_1)\\
&- P_{xtalk}(x_1-x_2,y_1-y_2) \G(x_2,y_2), 
\end{split}
\label{eq:crosstalkcorrection}
\end{equation}
where $\GG_{m}$ is the measured correlation function. Fig. \ref{fig:crosstalk} the anti-correlation peak of the far-field signal (a) is shown with corresponding estimated cross-talk probability map (b), which is point-symmetric with respect to $(\Delta x, \Delta y)=(0,0)$.

\section{Measurement results} \label{sec:Results}

\begin{figure}[htb]
	\begin{subfigure}[b]{0.45\linewidth}
		\includegraphics[width=0.8\linewidth]{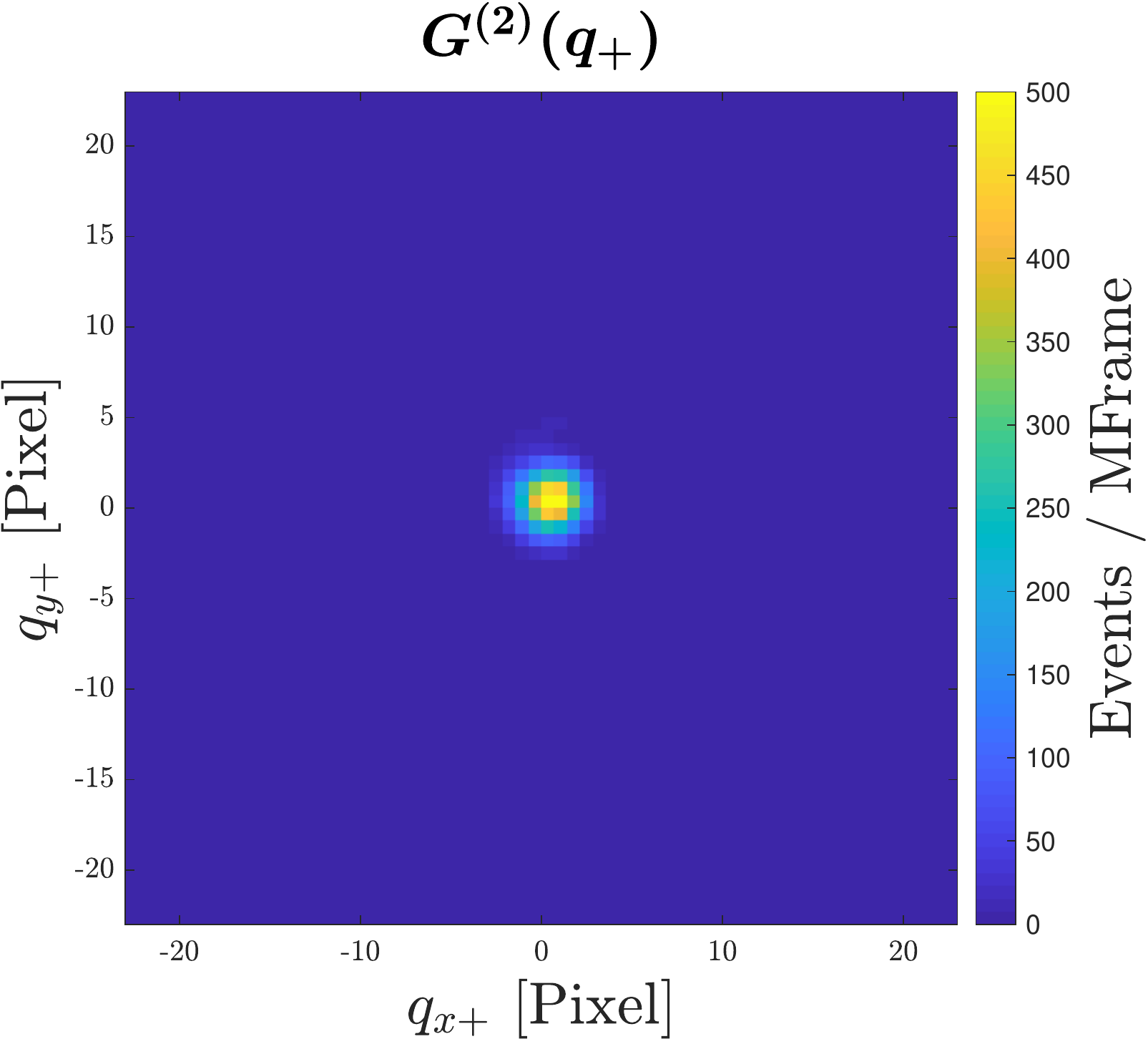}
		\subcaption{}
	\end{subfigure}
	\begin{subfigure}[b]{0.45\linewidth}
		\includegraphics[width=0.8\linewidth]{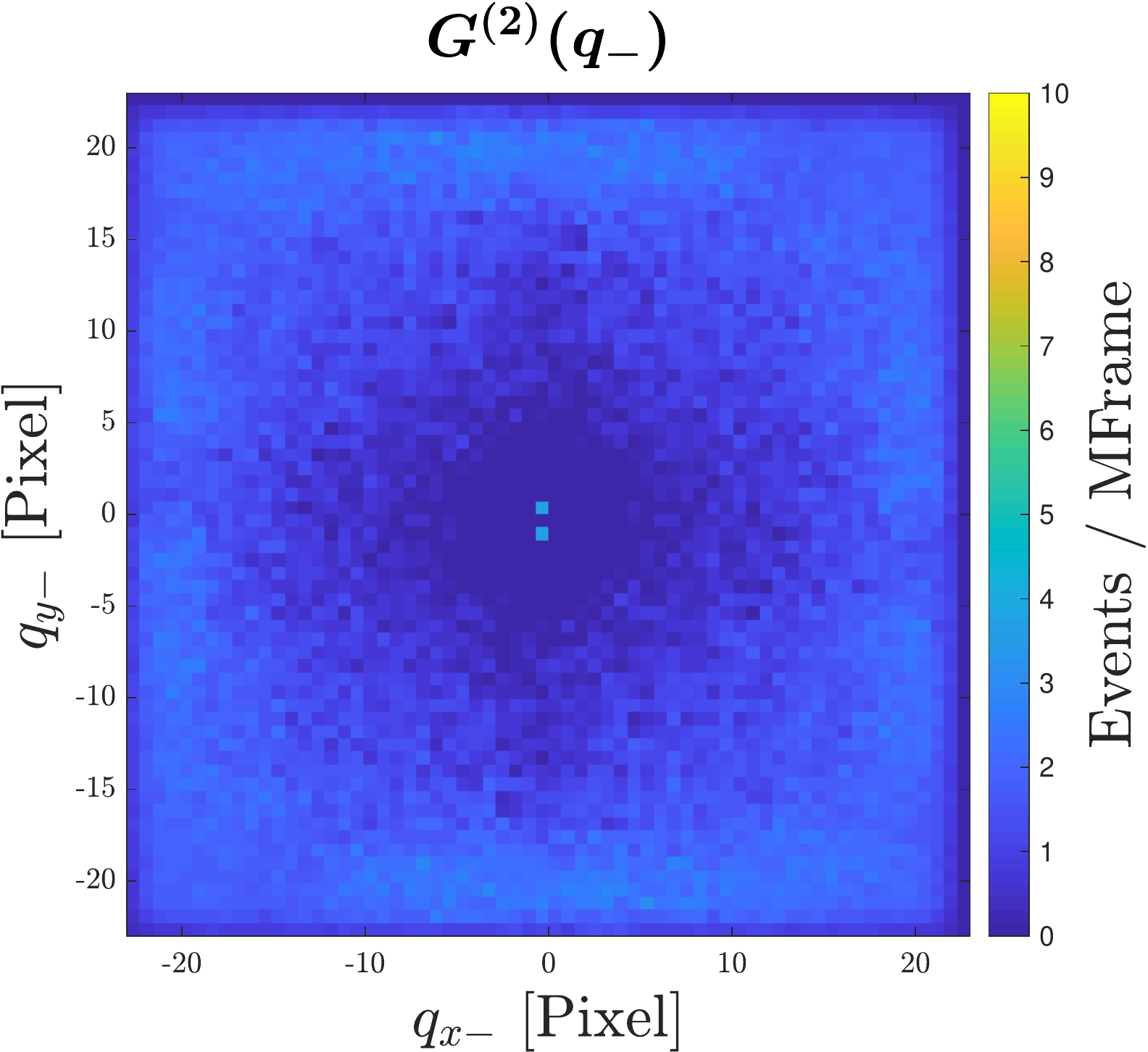}
		\subcaption{}
	\end{subfigure}
	\caption{Anti-correlation (a) and correlation (b) peaks of the second order far-field correlation. }
	\label{fig:g2_FF_plus_minus}
\end{figure}

The full correlation map results of far-field measurements are shown on Fig. \ref{fig:g2full_accidentals}. The same data can then be represented in the other coordinates previously introduced. Fig. \ref{fig:g2_FF_plus_minus} shows the anti-correlation (a) and correlation (b) peaks in the far-field after cross-talk removal.  As far-field measurements map the momentum space onto the sensor, the plot are labeled with the coordinates $\q_+$ and $\q_-$, but still in units of pixel. We clearly observe the presence of an anti-correlation peak and the almost disappearance of the correlation peak.

\begin{figure}[h]
	\begin{subfigure}[b]{0.45\linewidth}
		\includegraphics[width=0.8\linewidth]{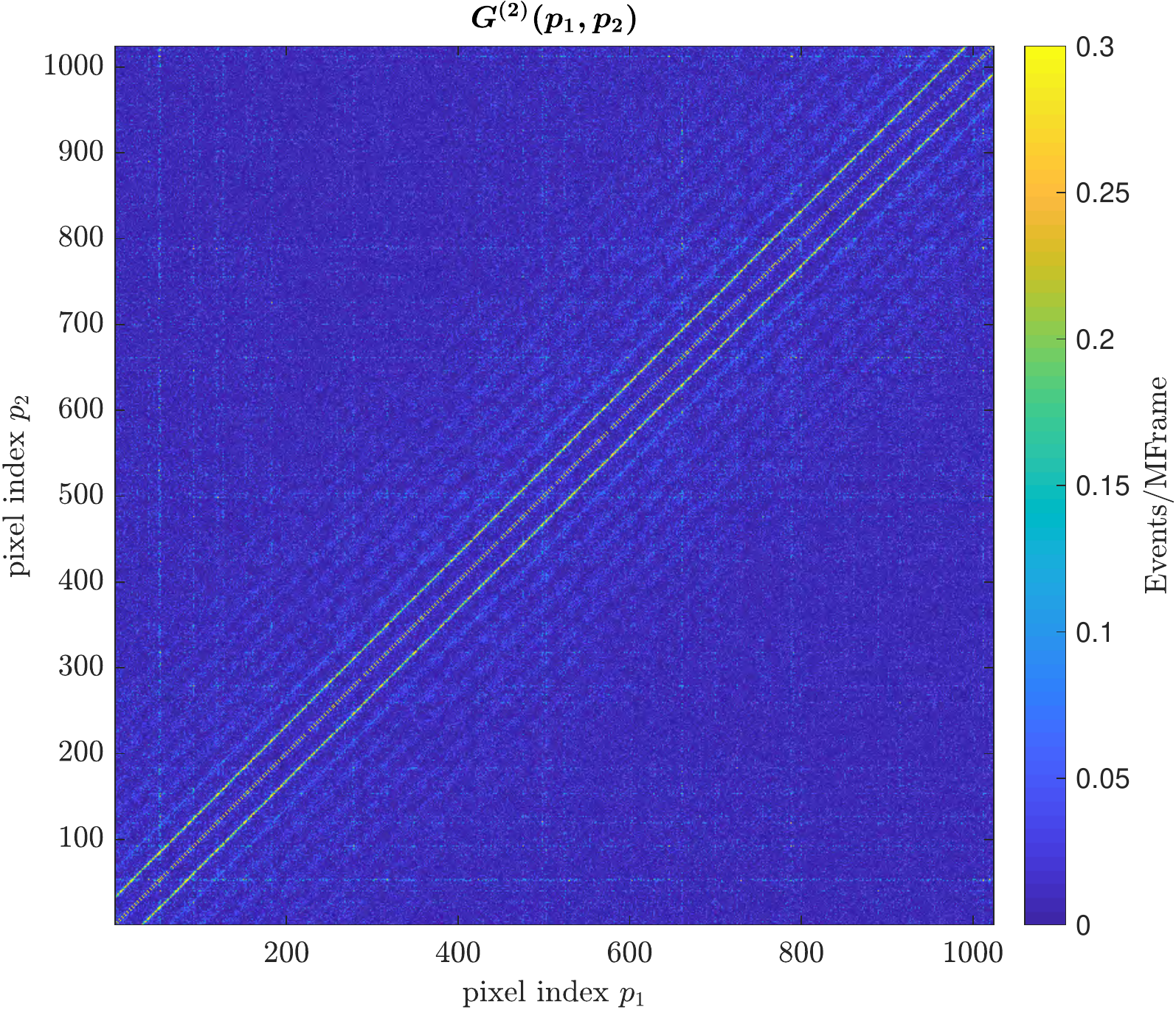}
		\subcaption{}
	\end{subfigure}
	\begin{subfigure}[b]{0.45\linewidth}
		\includegraphics[width=0.8\linewidth]{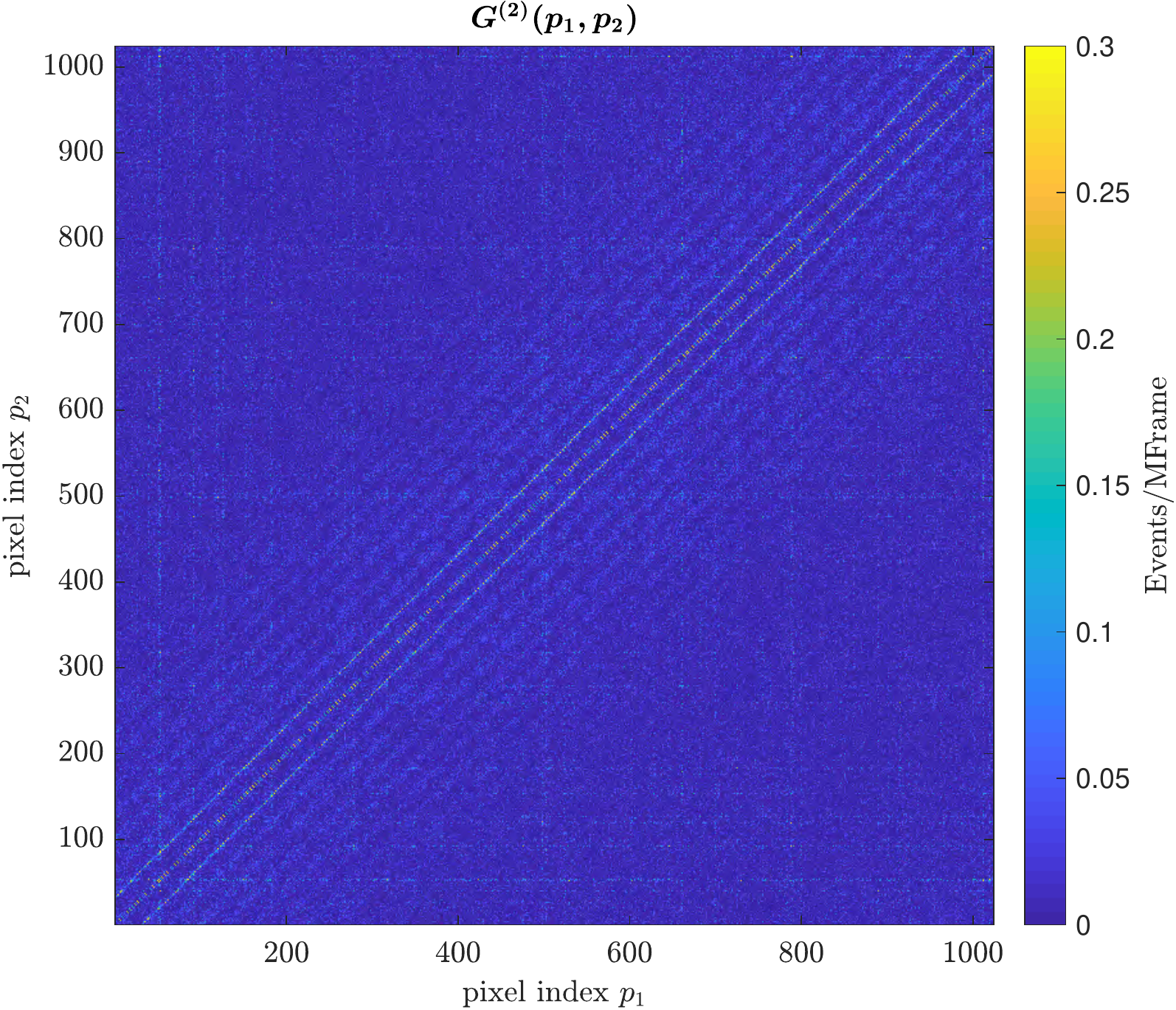}
		\subcaption{}
	\end{subfigure}
	\begin{subfigure}[b]{0.45\linewidth}
		\includegraphics[width=0.8\linewidth]{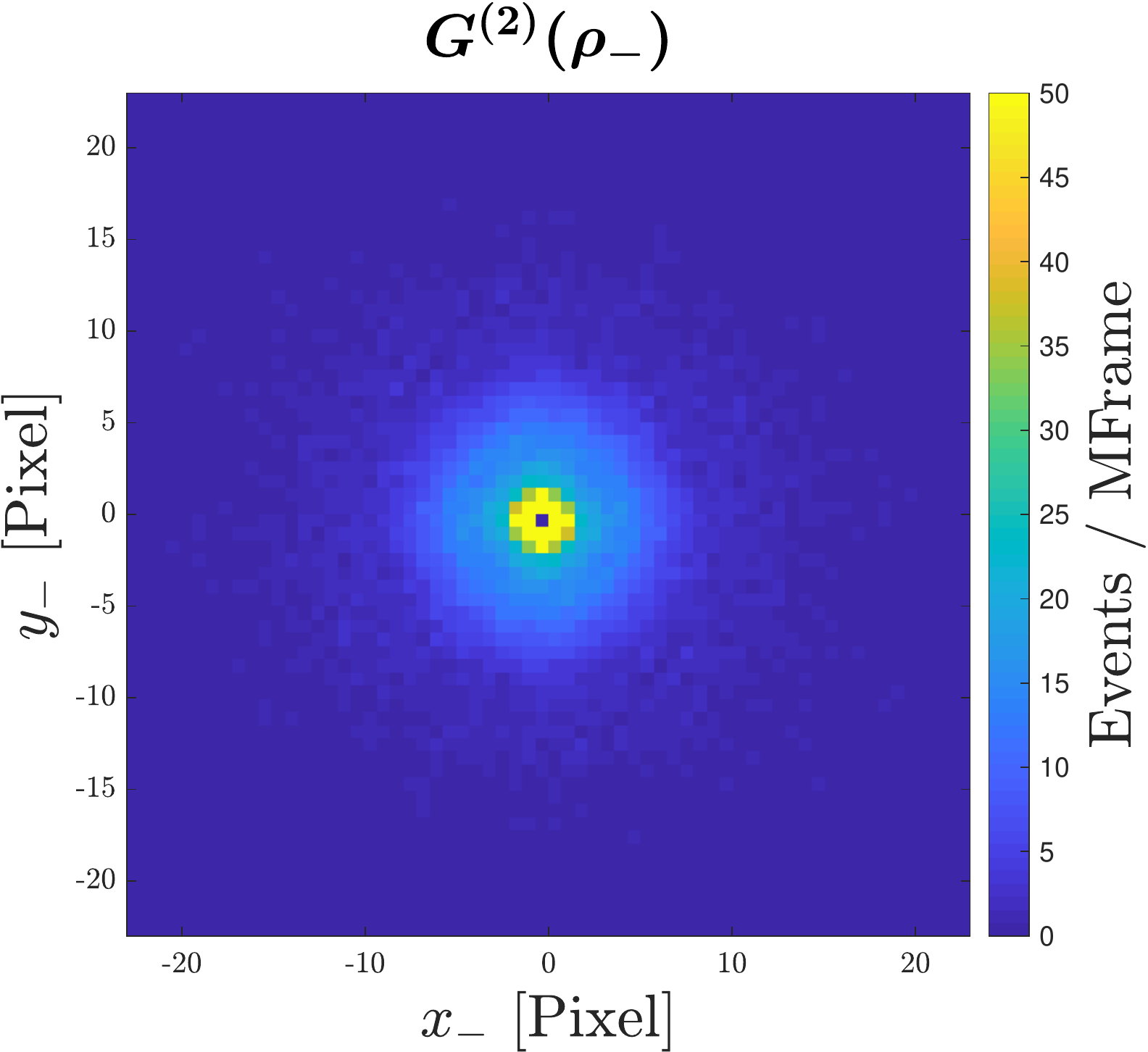}
		\subcaption{}
	\end{subfigure}
	\begin{subfigure}[b]{0.45\linewidth}
		\includegraphics[width=0.8\linewidth]{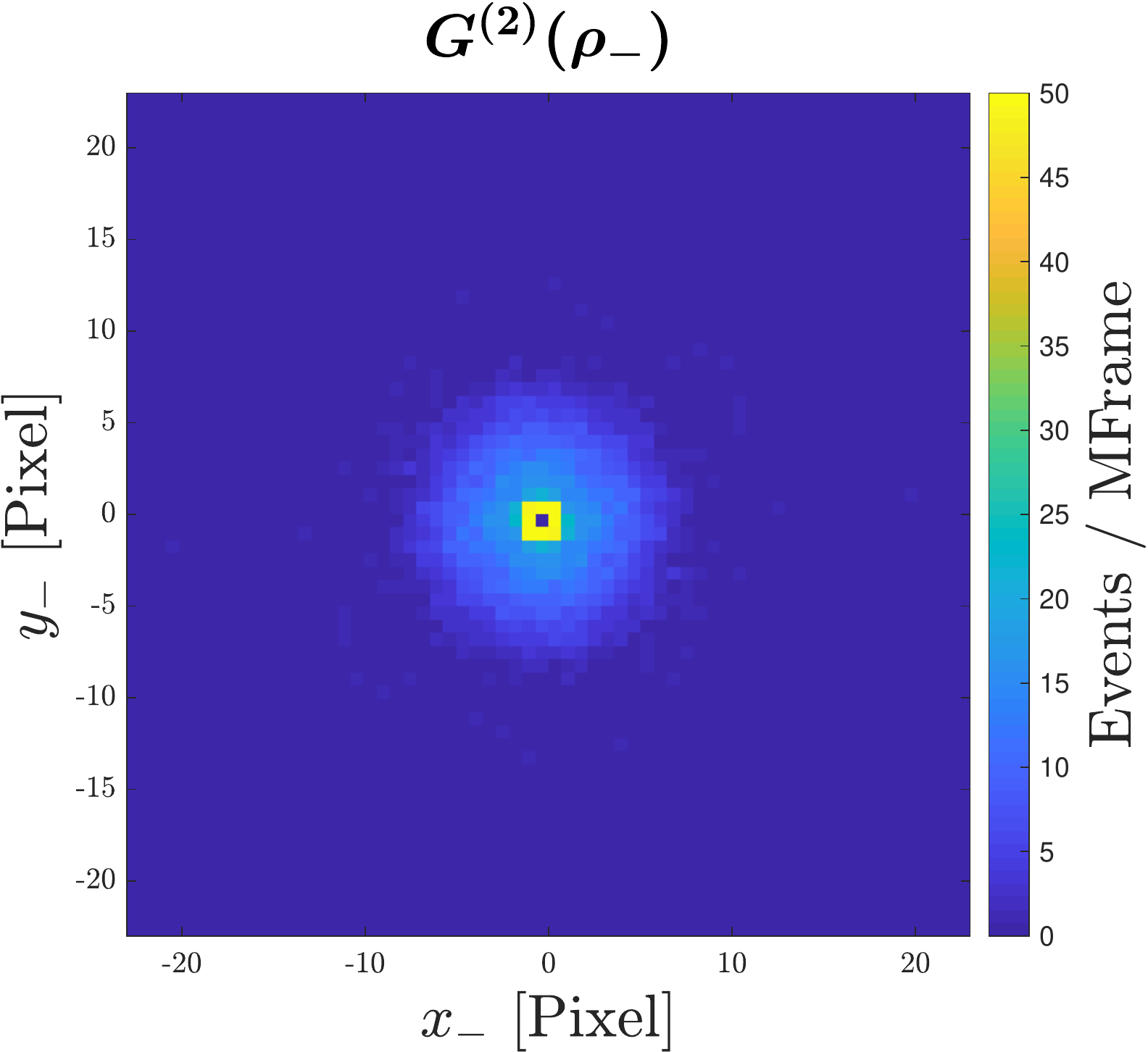}
		\subcaption{}
	\end{subfigure}
	\caption{Second order near-field correlation with accidentals removed, (a) Full correlation matrix  and (c) correlation peak. After cross-talk correction, (b) full correlation matrix and correlation peak (d).
	}
	\label{fig:g2_NF_beforeafterCT}
\end{figure}

The same processing can be applied for the near-field measurements. The full correlation matrix are shown in Fig. \ref{fig:g2_NF_beforeafterCT} without (a) and with (b) cross-talk correction. We observe that the cross-talk along the diagonal overlays the actual signal, even after correction. As the cross-talk contribution is of the same order as the signal, a correction is very sensitive to the estimation of the cross-talk probabilities. A full calibration at the pixel level will be performed in the future. This is why in the present work, the next neighbor correlations will not be taken into account in the quantitative estimation of the correlations. The partial suppression of the cross-talk can also be see on the correlation peaks of Fig. \ref{fig:g2_NF_beforeafterCT} (c) and (d), where the bright central peak is due to cross-talk, while the broad peak indicates correlation between photons. Note that the magnification factor has been chosen in order to obtain a correlation peak broader than the cross-talk.

\begin{figure}[h]
	\begin{subfigure}[b]{0.45\linewidth}
		\includegraphics[width=0.8\linewidth]{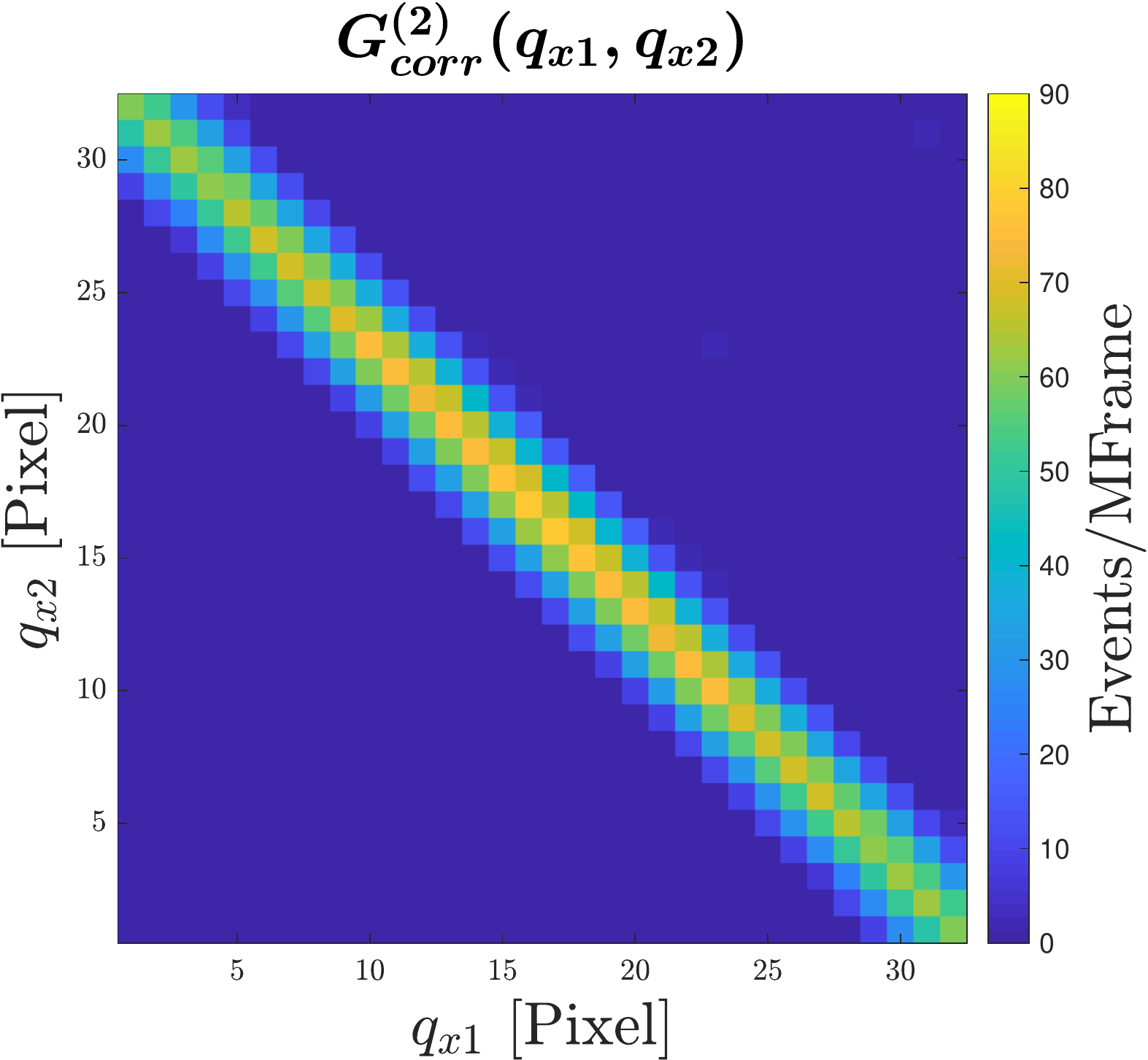}
		\subcaption{}
	\end{subfigure}
	\begin{subfigure}[b]{0.45\linewidth}
		\includegraphics[width=0.8\linewidth]{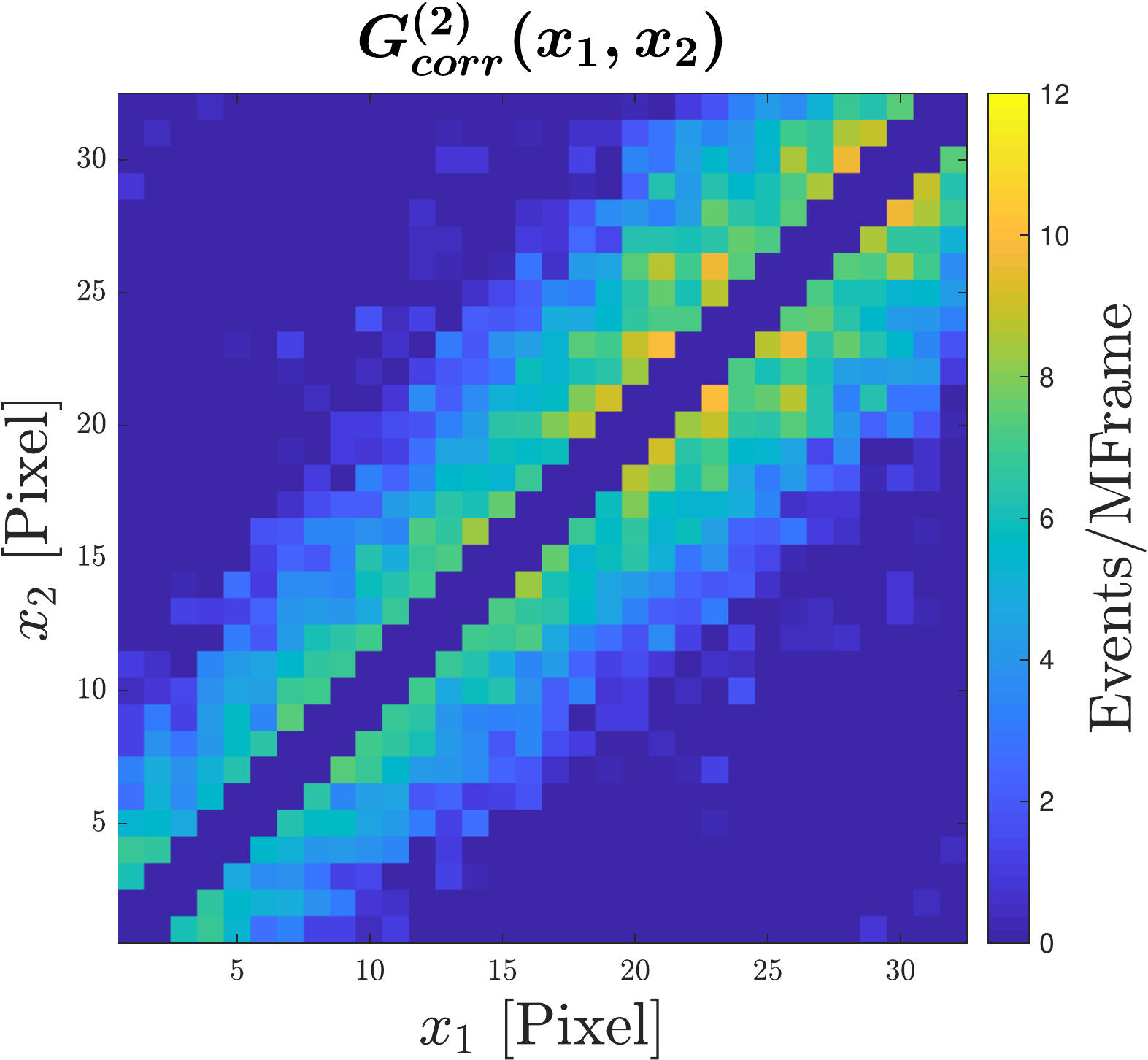}
		\subcaption{}
	\end{subfigure}
	\begin{subfigure}[b]{0.45\linewidth}
		\includegraphics[width=0.8\linewidth]{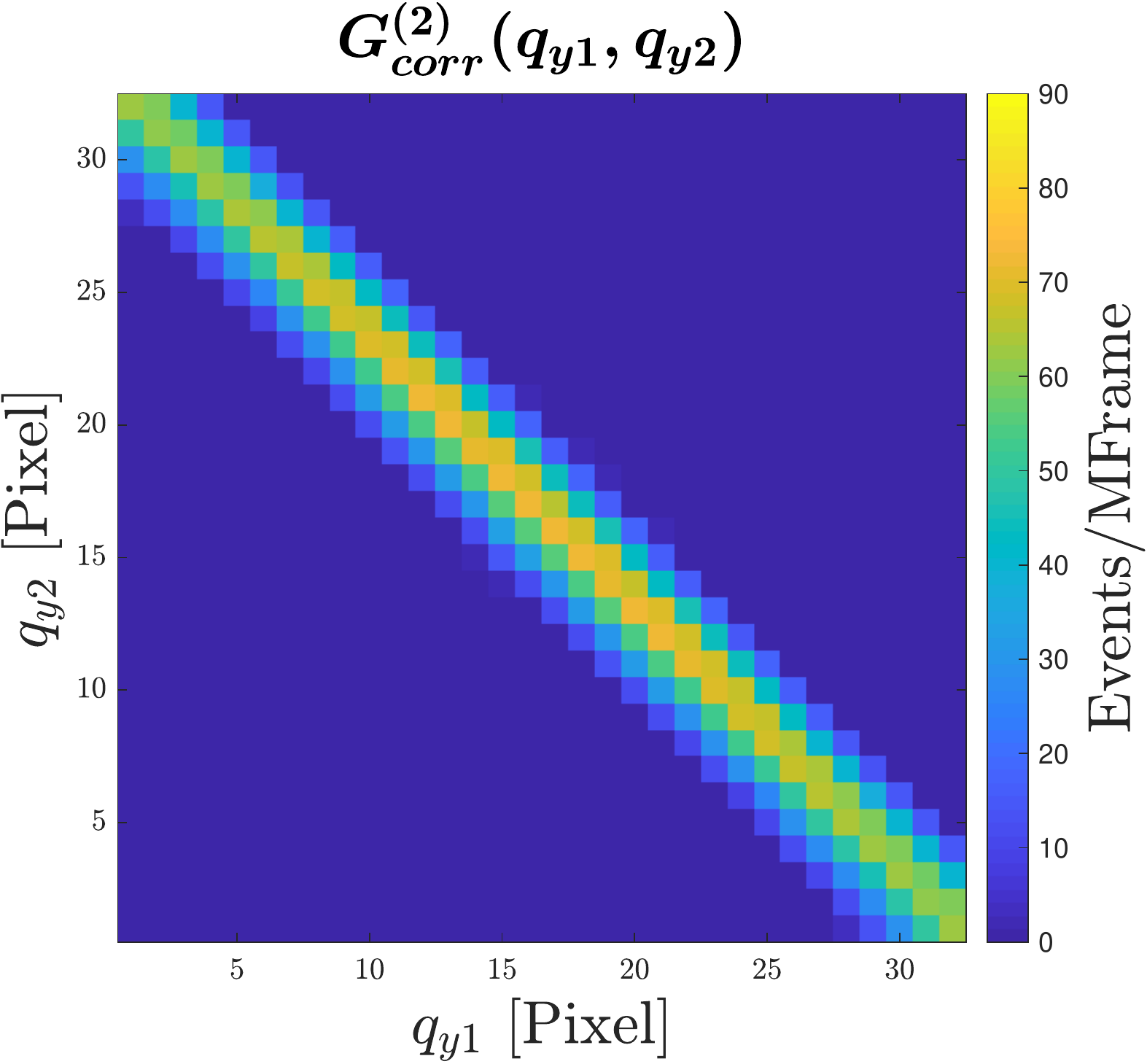}
		\subcaption{}
	\end{subfigure}
	\begin{subfigure}[b]{0.45\linewidth}
		\includegraphics[width=0.8\linewidth]{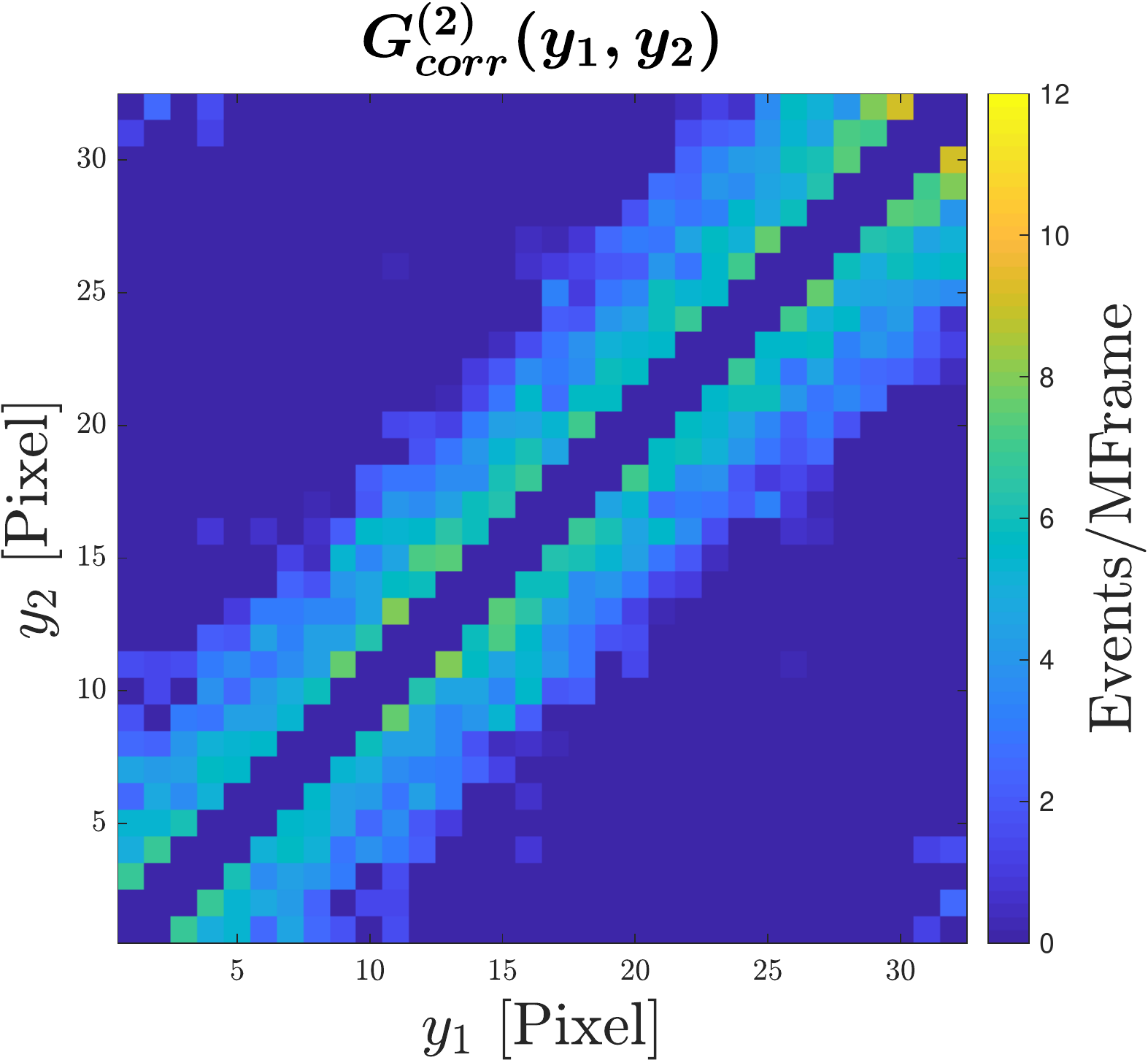}
		\subcaption{}
	\end{subfigure}
	\caption{Second order correlations in the far-field (a,c) and near-field (b, d) for the $x$ (a,b) and $y$ (c,d) coordinate.}
	\label{fig:data_x_y}
\end{figure}

Figures \ref{fig:data_x_y} show the second order correlations projected onto the $x$ and $y$ coordinates after removing cross-talk in the far- (a) and (c) and near-field (b) and (d). The pixel correlations mostly affected by cross-talk are set to 0. The anti-correlations in momentum and correlations in space are clearly visible. Those are the data further used to evaluate the EPR-type inequality.

While the identification of EPR-type correlations should be ideally done on the raw data, usually some additional assumption are used to correct for the imperfections of the sensor. Here, accidentals and cross-talk corrections are needed, as the detector sensitivity is limited at the photons wavelength. In a second step, the data from the second order correlations can be either directly numerically evaluated, or fitted with a model of the SPDC emission. The results of the various evaluations are summarized on table \ref{tab:resultsdpdx}.

\subsubsection{Numerical Evaluation of the EPR inequality}
First, we present in detail the numerical evaluation for the $x$ coordinates, i.e. $\GG(x_1,x_2)$ and $\GG(q_{x_1},q_{x_2})$. The treatment for the $y$ coordinate is identical.
First the conditional and unconditional probability density functions are estimated from the corrected experimental data
\begin{align}
	\mathcal{P}(x_1|x_2) &= \frac{\GG(x_1|x_2)}{\sum_{x_1} \GG(x_1|x_2)},\\
	\mathcal{P}(q_{x_1}|q_{x_2}) &= \frac{ \GG(q_{x_1}|q_{x_2})}{\sum_{q_{x_1}}\GG(q_{x1}|q_{x_2})},\\
	\mathcal{P}(x_2) &= \frac{\sum_{x_1} \GG(x_1,x_2)}{\sum_{x_1}\sum_{x_2} \GG(x_1,x_2)},\\
	\mathcal{P}(q_{x_2}) &= \frac{\sum_{q_{x_1}} \GG(q_{x_1},q_{x_2})}{\sum_{q_{x_1}}\sum_{q_{x_2}} \GG(q_{x_1},q_{x_2})}.
\end{align}

\begin{figure}[htb]
	\begin{subfigure}[b]{0.4\linewidth}
		\includegraphics[width=1\linewidth]{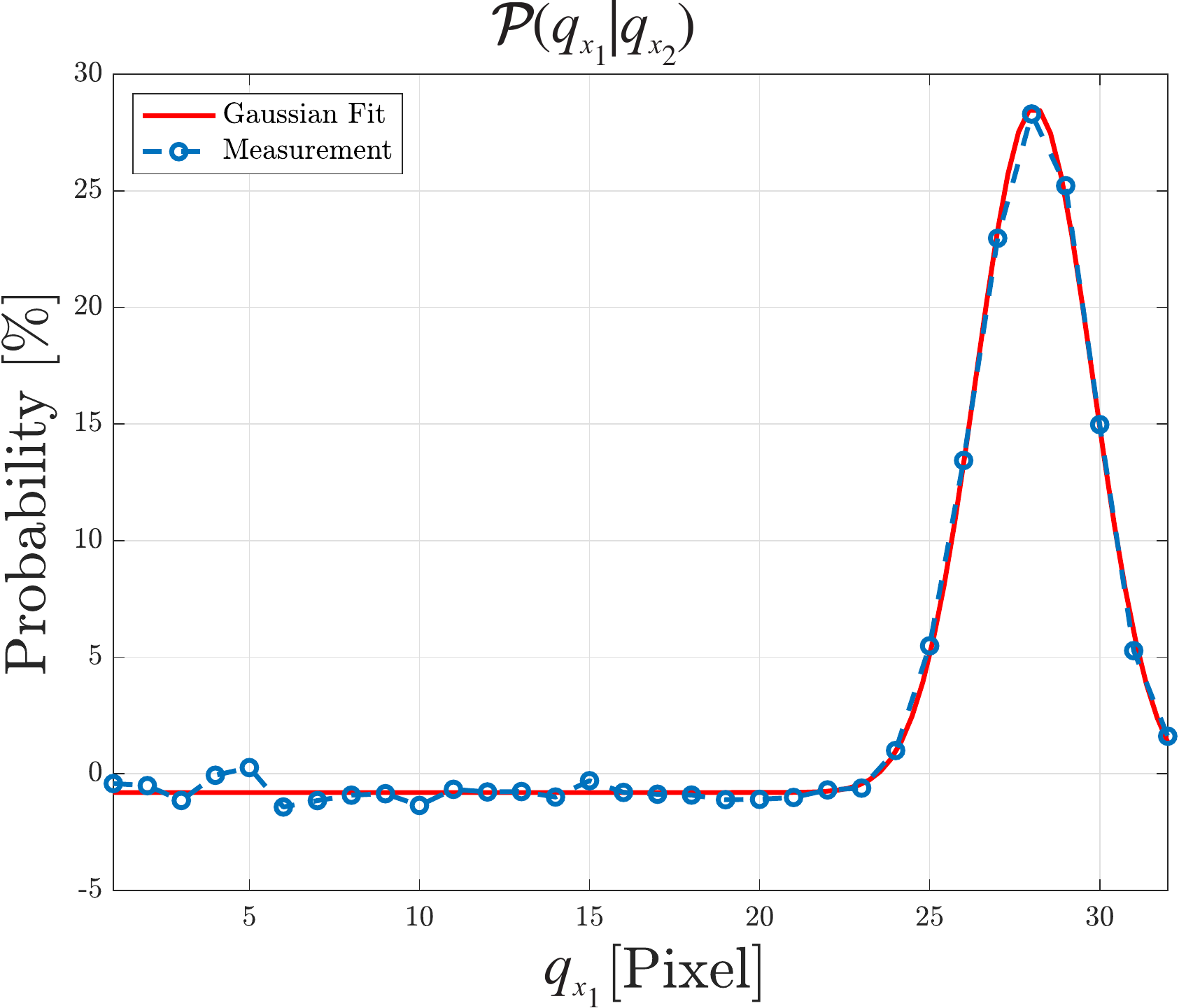}
		\subcaption{}
	\end{subfigure}
	\begin{subfigure}[b]{0.4\linewidth}
		\includegraphics[width=1\linewidth]{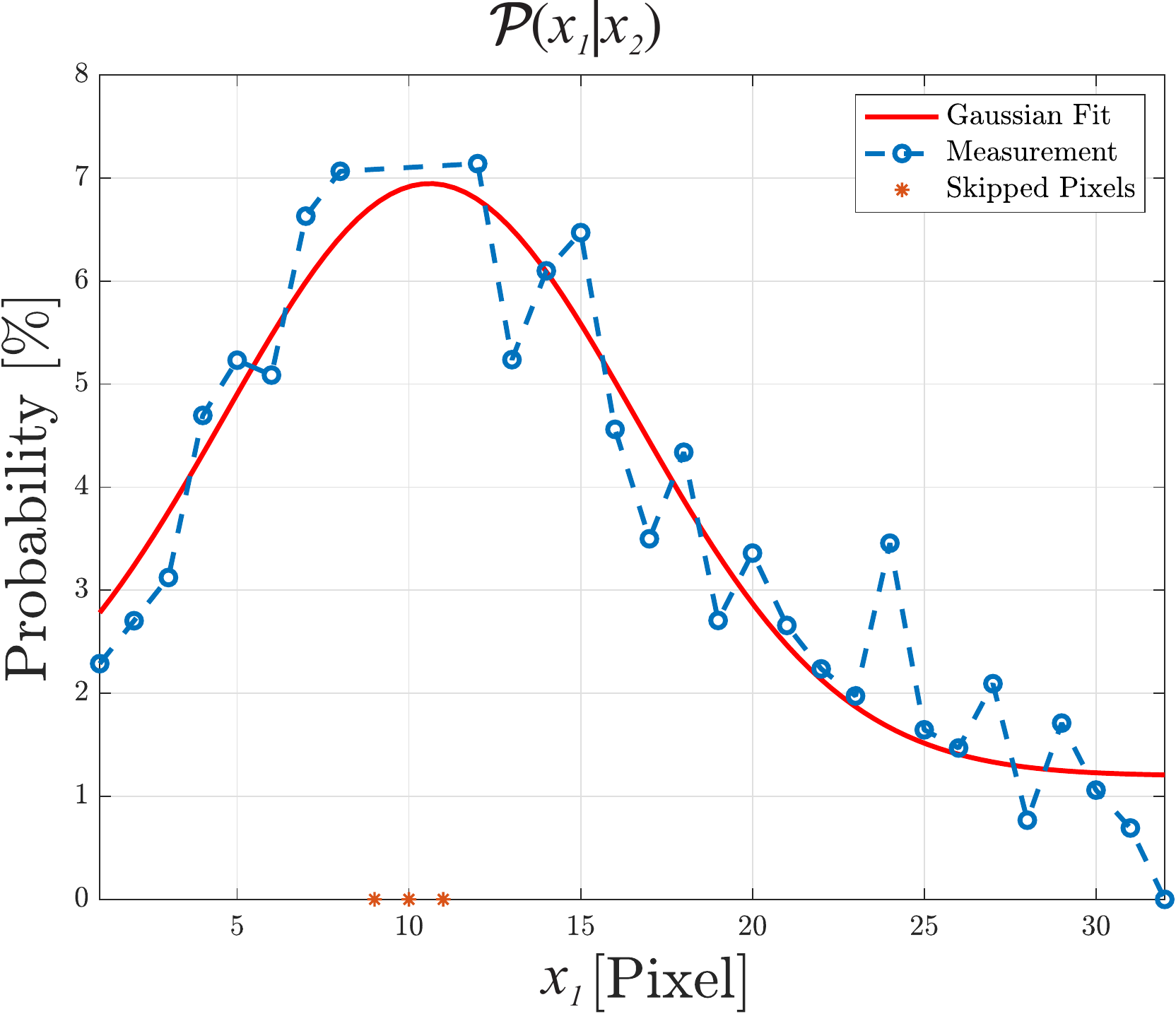}
		\subcaption{}
	\end{subfigure}
	\caption{Example of the measured conditional probabilities for the $x$ variable in the far-field $\mathcal{P}(q_{x_1}|q_{x_2})$ (a) and near-field $\mathcal{P}(x_{1}|x_{2})$ (b). The variance is either extracted numerically or using a Gaussian fit. The pixels affected from cross-talk are skipped (indicated by the red asterisks)}
	\label{fig:probabilities_conditioned}
\end{figure}

Examples of experimentally obtained probability distributions are shown on Fig. \ref{fig:probabilities_conditioned}. Next, expectation values are given by
\begin{align}
	\mu_{x1} &= \sum_{x_1} x_1\  \mathcal{P}(x_1|x_2),\\
	\mu_{q_{x1}} &= \sum_{q_{x_1}} q_{x_1}\  \mathcal{P}(q_{x_1}|q_{x_2}),
\end{align}
and the conditional variances are
\begin{align}
	\Delta^2(x_1|x_2) &= \sum_{x_1}(x_1-\mu_{x1})^2\  \mathcal{P}(x_1|x_2),\\
	\Delta^2(q_{x_1}|q_{x_2}) &= \sum_{q_{x_1}}(q_{x_1}-\mu_{q_{x1}})^2\  \mathcal{P}(q_{x_1}|q_{x_2}).
\end{align}
Finally, the minimum inferred variances are computed
\begin{align}
	\Delta^2_{min}(x_1|x_2) &= \sum_{x_2} \mathcal{P}(x_2) \ \Delta^2(x_1|x_2),\\
	\Delta^2_{min}(q_{x_1}|q_{x_2}) &= \sum_{q_{x_2}}  \mathcal{P}(q_{x_2}) \ \Delta^2(q_{x_1}|q_{x_2}).
\end{align}
We introduce the quantity
\begin{align}
	V_{min}^{(x)} \equiv \Delta^2_{min}(x_1|x_2)\cdot \Delta^2_{min}(q_{x_1}|q_{x_2}),
\end{align}
and EPR-type correlations are identified by violating the inequality
\begin{equation}
V_{min}^{(x)}  > \frac{1}{4}.
\end{equation}

The same procedure is applied for the $y$ coordinate and the result is denoted as 
\begin{align}
	V_{min}^{(y)} \equiv \Delta^2_{min}(y_1|y_2)\cdot \Delta^2_{min}(q_{y_1}|q_{y_2}).
\end{align}

The direct numerical evaluation of $V_{min}$ from the corrected measured data is sensitive to noise and to the accidental and cross-talk corrections. The measured conditional probability distributions take non-zero values even far away from their expected peak value, as can be seen in Fig. \ref{fig:probabilities_conditioned}. This effect shifts the expectation value $\mu_{x_1,q_{x1}}$ away from the true one and also increases the conditional variances. The second challenge arises from the cross-talk. Setting to $0$ the pixels mostly affected in the near-field correlations increases the variance of the conditioned probability distributions. As a consequence the numerically determined values of $V_{min}$ given in Table \ref{tab:resultsdpdx} are overestimated

\begin{figure}[htb]
	\begin{subfigure}[b]{0.4\linewidth}
		\includegraphics[width=1\linewidth]{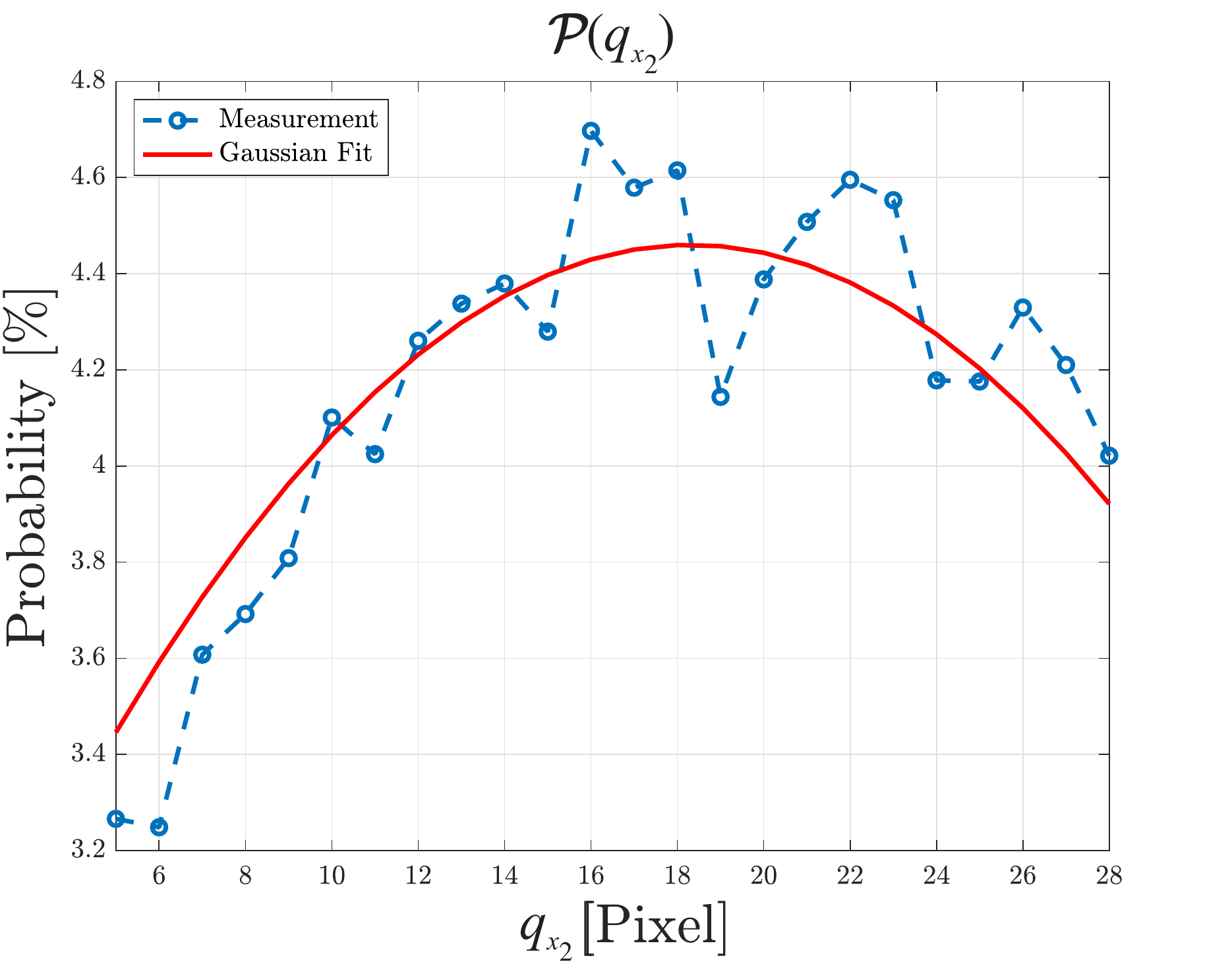}
		\subcaption{}
	\end{subfigure}
	\begin{subfigure}[b]{0.4\linewidth}
		\includegraphics[width=1\linewidth]{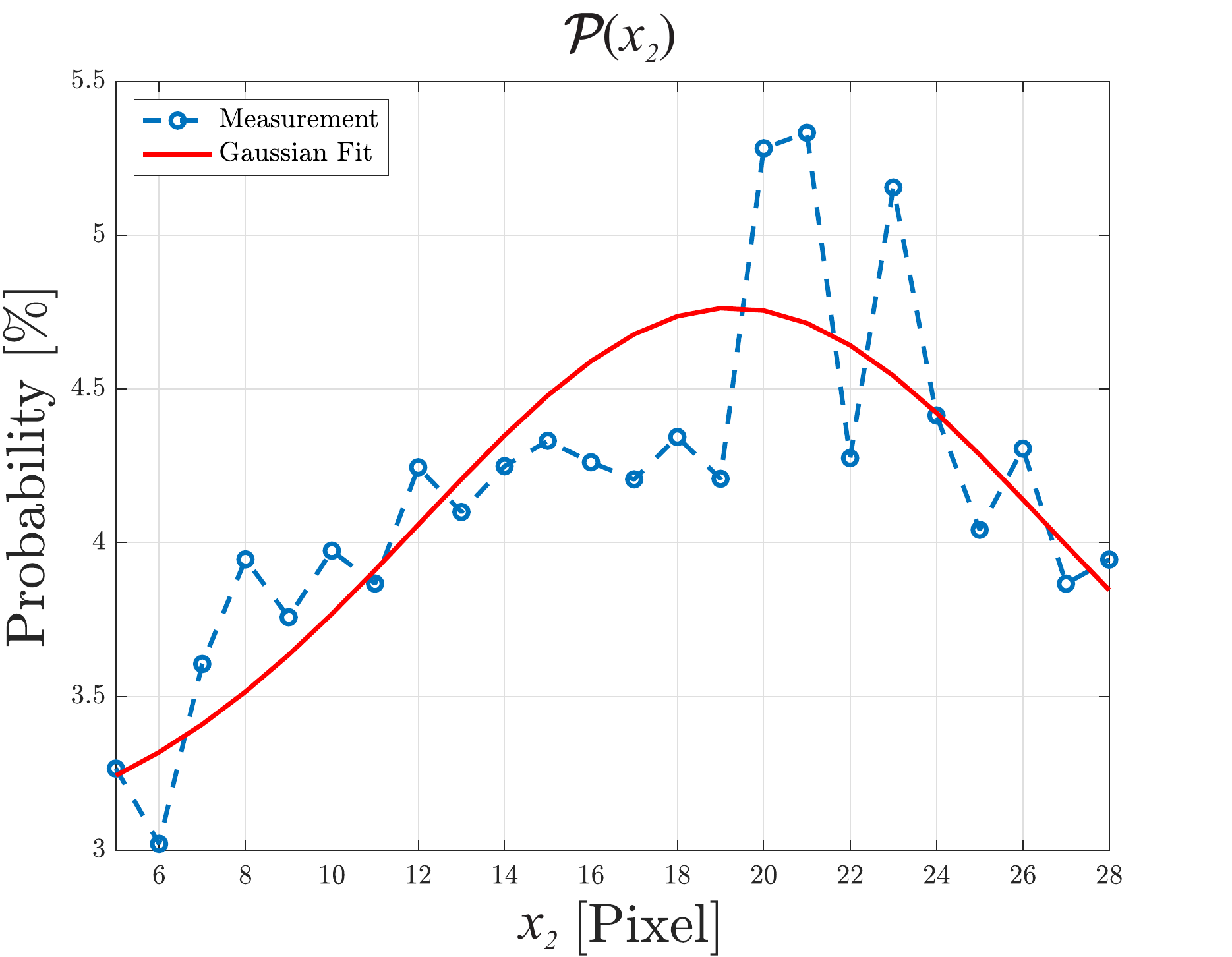}
		\subcaption{}
	\end{subfigure}
	\caption{Example of the measured unconditional probabilities for the $x$ variable in the far-field (a) and near-field (b).}
	\label{fig:probabilities_unconditioned}
\end{figure}
\subsubsection{1D Gaussian Fitting}
An alternative approach to direct numerical evaluation is to constrain the data into a set of "reasonable" data. The shape of the JSA from SPDC emission can be, in some conditions, approximated by Gaussian functions. Therefore fitting Gaussian functions to the data is a way to to obtain $P(q_{x_2})$, $P(x_{2})$, $P(q_{x_1}|q_{x_2})$ and $P(x_{1}|x_{2})$. Figures \ref{fig:probabilities_conditioned} and \ref{fig:probabilities_unconditioned} show example of fitted data. One can then extract the variances  directly from these fitted probability density functions and calculate the EPR-criterion according to the definition of Eqs. \eqref{eq:epr_minimuminferredvariance_x} and \eqref{eq:EPRInequality}. The results are shown in the second line of Table \ref{tab:resultsdpdx} and are in agreement with the expected values from the experimental parameters.

\subsubsection{2D Gaussian Fitting}
\begin{figure}[htb]
	\begin{subfigure}[b]{0.45\linewidth}
		\includegraphics[width=1\linewidth]{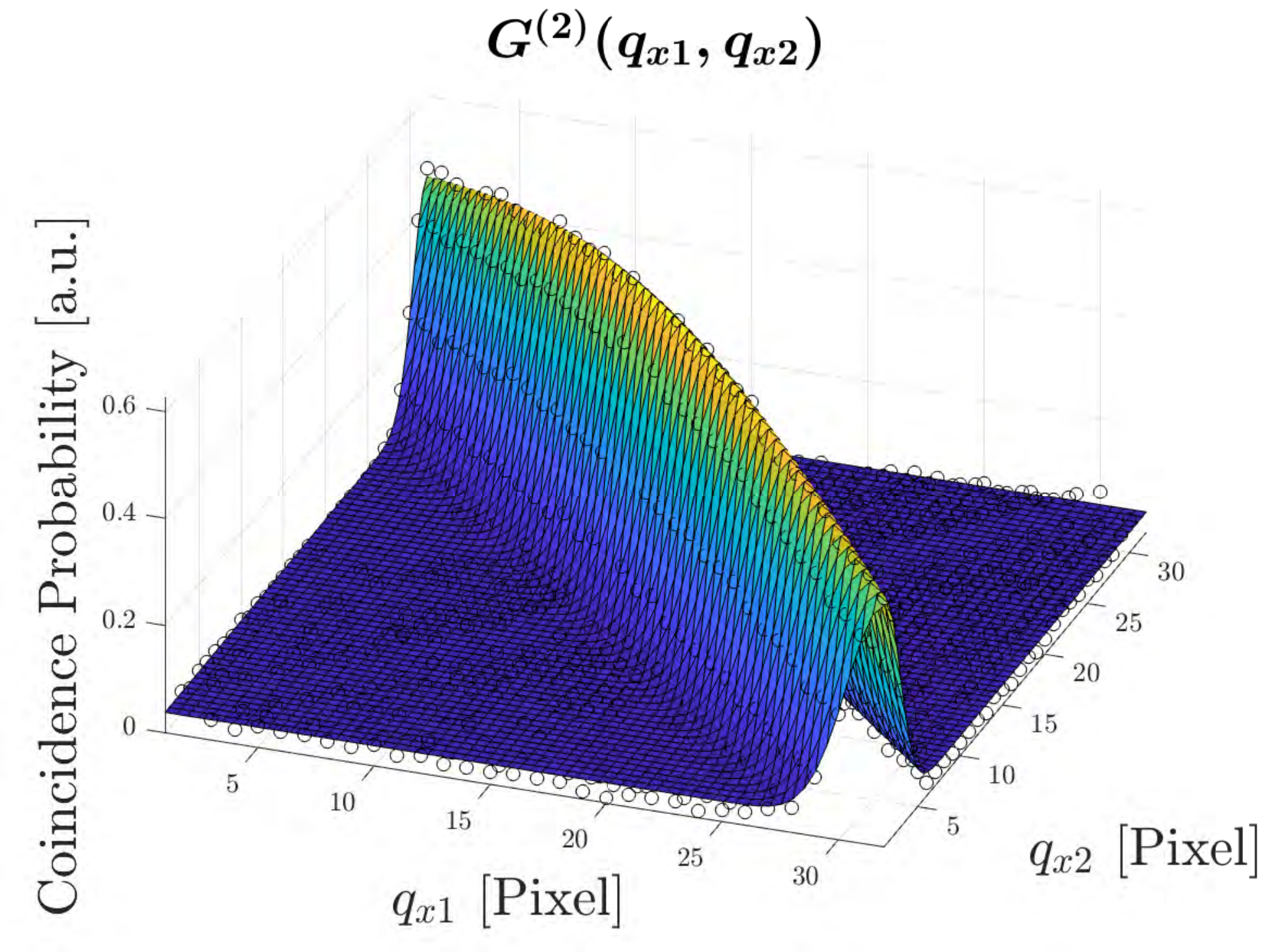}
		\subcaption{}
	\end{subfigure}
	\begin{subfigure}[b]{0.45\linewidth}
		\includegraphics[width=1\linewidth]{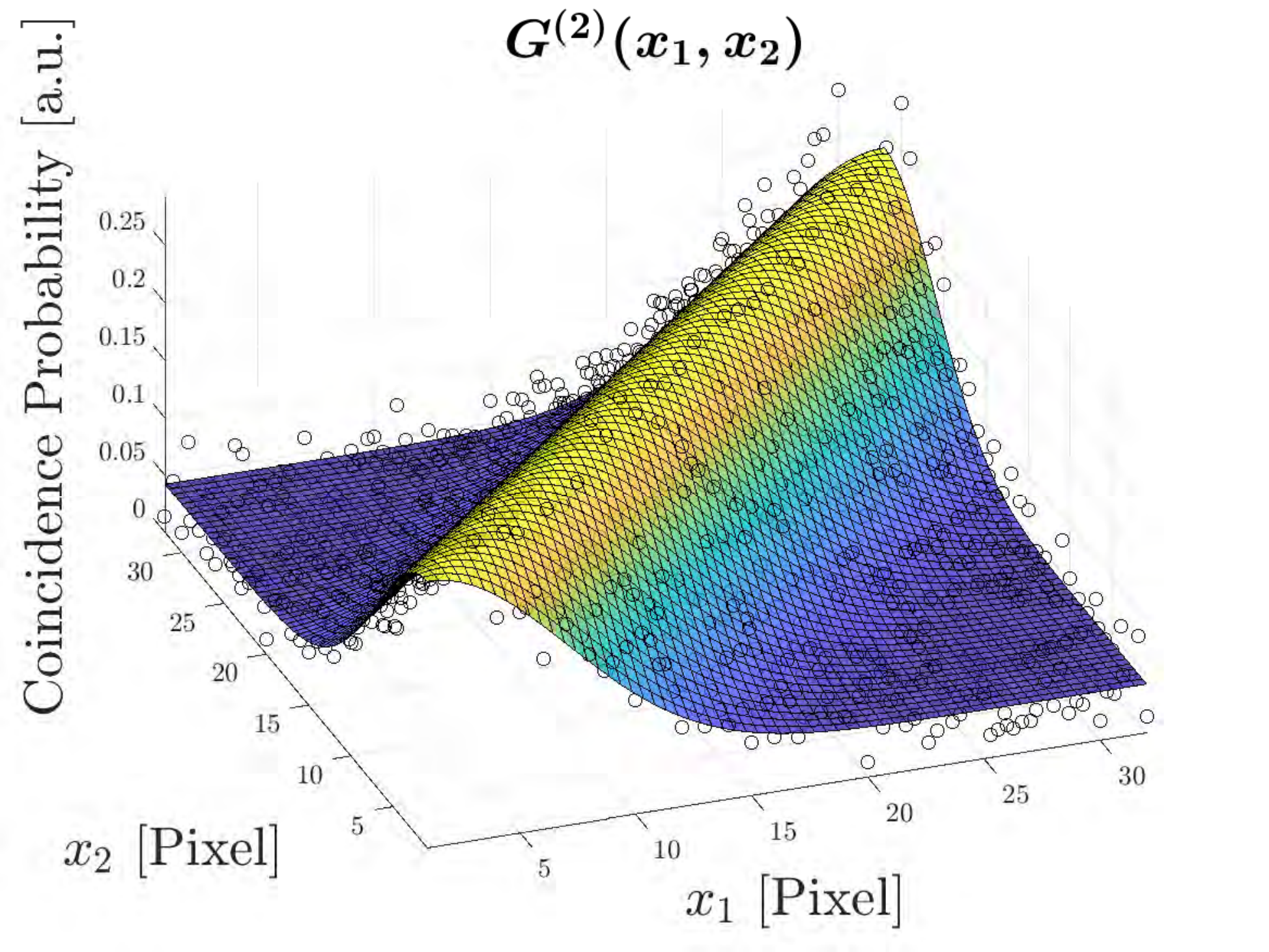}
		\subcaption{}
	\end{subfigure}
	\caption{The second order correlations in the far-field (a) and near-field (b) are directly fitted with 2D Gaussian distributions and the variances are extracted from the fits.}
	\label{fig:g2_doublegaussfitting}
\end{figure}

Instead of fitting multiple Gaussians by \text{slicing} the data sets $\GG$, one can also directly fit 2D Gaussians over $\GG(x_1,x_2)$ and $\GG(q_{x_1},q_{x_2})$ (and similarly for $y$), as shown on Fig. \ref{fig:g2_doublegaussfitting}. Note that we only assume the correlations to be well-approximated by 2D Gaussians with variances $\Delta^2_{x+,x-}$ and $\Delta^2_{qy+,qx-}$, but we don't impose the quantum state to follow the double gaussian model \cite{Fedorov2009}. The conditional variances are then related to the fitting parameters by \cite{Schneeloch2016}
\begin{align}
	\Delta^2(x_1|x_2) &= \frac{2 \Delta^2_{x+}\Delta^2_{x-}}{\Delta^2_{x+}+\Delta^2_{x-}},\\
	\Delta^2(q_{x1}|q_{x2}) &= \frac{2 \Delta^2_{qx+}\Delta^2_{qx-}}{\Delta^2_{qx+}+\Delta^2_{qx-}}.
\end{align}

\subsubsection{(Anti-) Correlation Peaks}
\begin{figure}[hb]
	\begin{subfigure}[b]{0.3\linewidth}
		\includegraphics[width=1\linewidth]{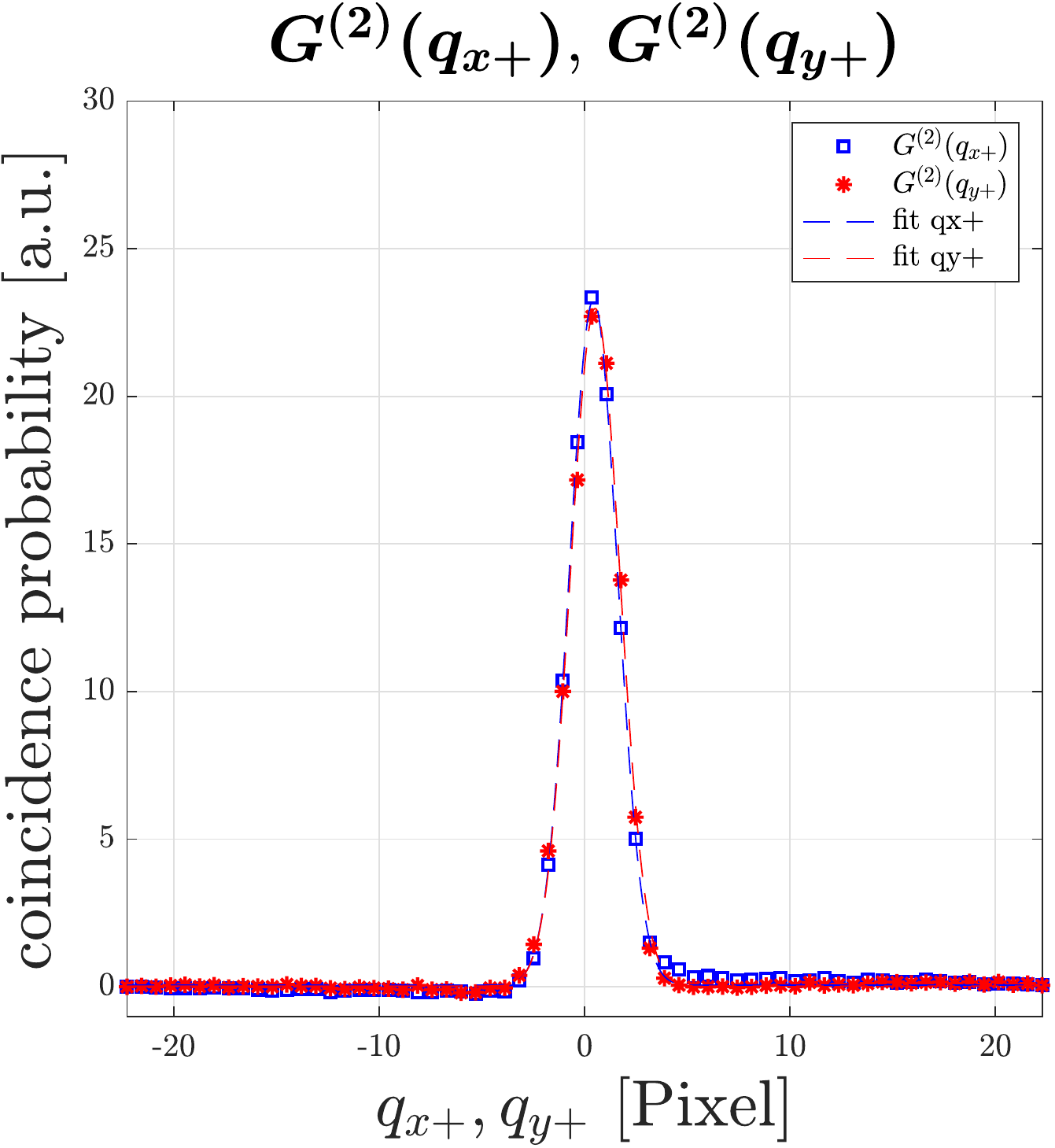}
		\subcaption{}
	\end{subfigure}
	\begin{subfigure}[b]{0.3\linewidth}
		\includegraphics[width=1\linewidth]{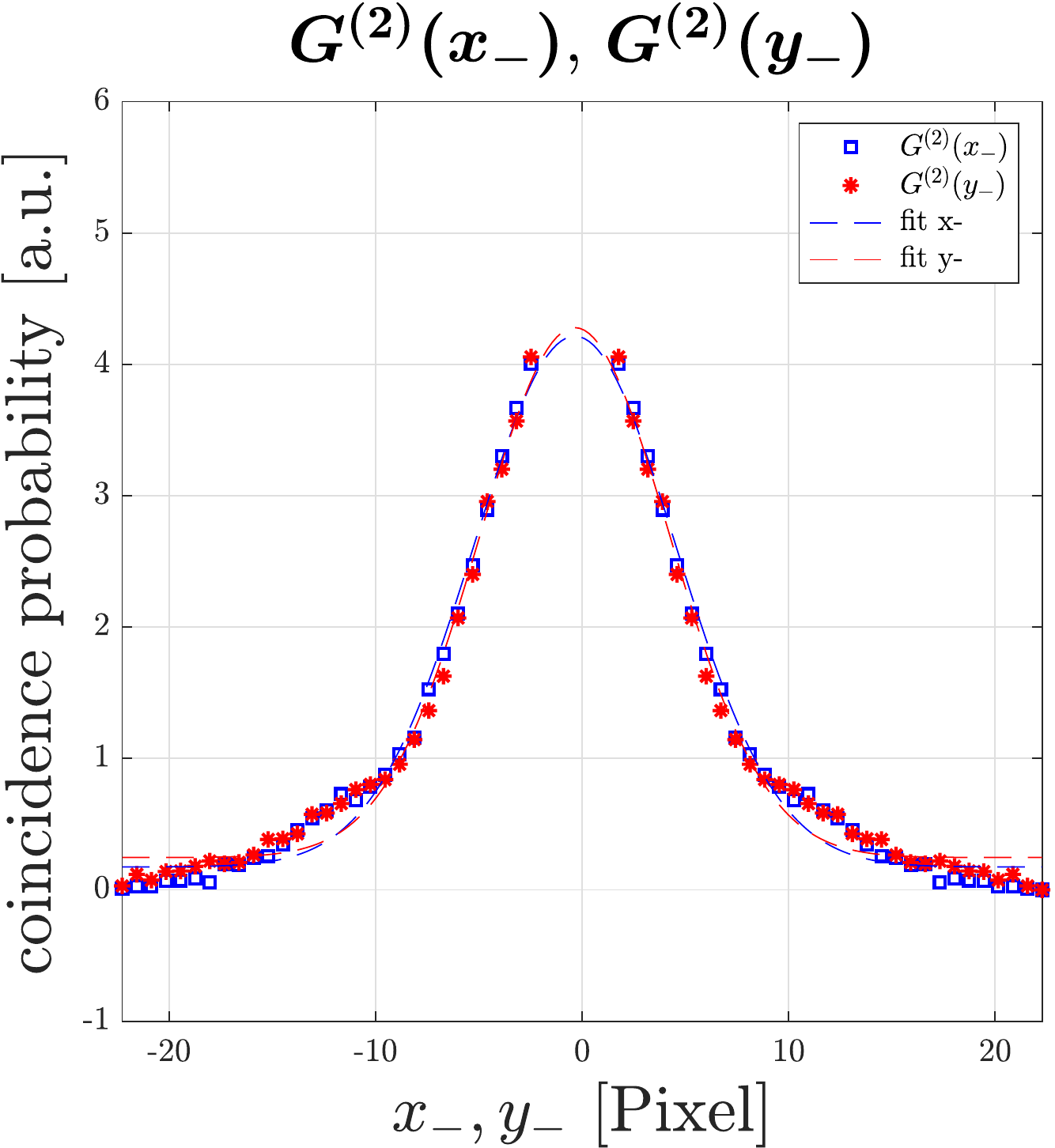}
		\subcaption{}
	\end{subfigure}
	\caption{Anti-correlations peaks (a) of the second order correlations in the far-field and correlation peaks (b) from near-field measurements alon $x$ and $y$ directions.}
	\label{fig:data_peaks}
\end{figure}

The projections onto $x$ and $y$ of the (anti-) correlation peaks $\GG(\q_+)$ and $\GG(\ro_-)$ are shown on Fig. \ref{fig:data_peaks}. The width of the fitted Gaussian are respectively $\sigma_{x+},\sigma_{y+}$ and $\sigma_{qx+},\sigma_{qy+}$. They can be related to the conditional variances if we assume the correlations to appear along the $+$ and $-$ coordinates. Next we require the correlations in the near-field to be expanded very far along the $+$ coordinate (i.e. $\sigma_{x+},\sigma_{y+} \to \infty$) and similarly in the far-field for the $-$ coordinate ($\sigma_{qx+},\sigma_{qy+} \to \infty$). We can then estimate the unconditioned variances from $\GG(x_-)$ and $\GG(q_{x+})$ (see Fig. \ref{fig:data_peaks}). The conditional variances (or its square roots) are then given by
\begin{align}
	\Delta(x_1|x_2) &\approx \sqrt{2} \sigma_{x-},\\
	\Delta(q_{x1}|q_{x2}) &\approx \sqrt{2}\sigma_{qx+}
\end{align}
and similarly for the $y$ coordinate.

\begin{table}[h]
	\centering
	\caption{Evaluation of the conditioned variances from second order correlation measurements}
	\label{tab:resultsdpdx}
	\begin{ruledtabular}\begin{tabular}{l|c|c|c|c|c|c|}  
			
			& $\Delta_{min}\left(q_{x1}|q_{x2}\right)$ & $\Delta_{min}\left(q_{y1}|q_{y2}\right)$ & $\Delta_{min}\left(x_1|x_2\right)$ & $\Delta_{min}\left(y_1|y_2\right)$ & $V_{min}^{(x)}$ & $V_{min}^{(y)}$\\
			& $[\u{}{\per\milli\meter}]$ & $[\u{}{\per\milli\meter}]$ & $[\u{}{\micro\meter}]$ & $[\u{}{\micro\meter}]$ & &\\
			
			\hline 
			\textit{numerical} & \u{6.3}{}    &\u{6.6}{} &\u{37.2}{}&\u{36.2}{} & \u{5.5e-2}{}    &\u{5.8e-2}{} \Tstrut\\
			\textit{1D Gaussians} & \u{3.9}{}    &\u{4.1}{} &\u{32.8}{}&\u{30.5}{} & \u{1.6e-2}{}    &\u{1.5e-2}{} \\
			\textit{$2$D Gaussian} & \u{3.9}{} &\u{4.1}{} &\u{34.4}{} 	&\u{31.1}{} & \u{1.8e-2}{}    &\u{1.6e-2}{} \\
			\textit{(anti-)correlation peak} & \u{3.8}{} &\u{4.0}{} &\u{34.3}{}&\u{30.3}{} & \u{1.7e-2}{} &\u{1.4e-2}{} \\
			\hline		
			\textit{expected values from source parameters} & \u{4.0}{} &\u{3.4}{} &\u{37.3}{}&\u{37.3}{} & \u{2.2e-2}{} &\u{1.6e-2}{} \Tstrut\\
	\end{tabular}\end{ruledtabular}
\end{table}%

\section{Discussion}
We make use of the capacity of a recently developed 32x32 pixels CMOS SPAD array sensor to time-tag each individual photons with high temporal resolution in order to measure the emission of spatially entangled photon pairs from SPDC. Anti-correlations in the far-field and correlations in the near-field are observed. From the experimentally measured probability distributions, the violation of an EPR-type inequality is demonstrated. Interestingly the inequality is violated even without fitting the data with a specific model. When assuming Gaussian shape for the probability distributions, the obtained value for the variances are in good agreement with the expected ones derived from the parameters of the photon source. Those results demonstrate the ability of CMOS SPAD arrays to effectively measure simultaneously spatial and temporal correlations between photons. They are a promising tool for practical implementations of quantum imaging schemes, that will require further developments towards higher number of pixels.

\begin{acknowledgments}
This research was supported by the EU project Horizon-2020 SUPERTWIN id.686731 and received funding from the European Union’s Horizon 2020 research and innovation programme under grant agreement No 852045.
\end{acknowledgments}
\bibliographystyle{nature}


\end{document}